\def\,{\thinspace}
\def \ppm{$\pm$}
\def\kms    {\ifmmode{{\rm \ts km\ts s}^{-1}}\else{\ts km\ts s$^{-1}$}\fi}
\def\kmspc{\kms\,pc$^{-1}$}

\def\sgra{Sgr\,A$^*$}
\def\etal   {\rm et\ts al. }
\def\msun {M$_\odot$}

\documentclass[oldversion]{aa}

\usepackage{epsfig}
\usepackage{lscape}
\usepackage{longtable}
\usepackage{multirow}
\usepackage{appendix}
\usepackage{lineno}

\usepackage{txfonts}
\begin{document}

   \title{The thermal state of molecular clouds in the Galactic center:
evidence for non-photon-driven heating}
   \author{Y. Ao \inst{1,2}\thanks{email: ypao@mpifr-bonn.mpg.de}, 
   C. Henkel \inst{1,3}, K. M. Menten\inst{1}, M. A. Requena-Torres\inst{1}, T. Stanke\inst{4}, R. Mauersberger\inst{5}, 
   S. Aalto\inst{6}, S. M\"{u}hle \inst{7}, and J. Mangum \inst{8}}
   \institute{
   Max-Planck-Institut f{\"u}r Radioastronomie, Auf dem H\"{u}gel 69, D53121 Bonn, Germany
   \and
   Purple Mountain Observatory, Chinese Academy of Sciences, Nanjing 210008, China
   \and
   Astron. Dept., King Abdulaziz University, P.O. Box 80203, Jeddah, Saudi Arabia
   \and
   ESO, Karl-Schwarzschild Strasse 2, 85748 Garching bei M{\"u}nchen, Germany
   \and
   Joint ALMA Observatory, Av. Alonso de C$\acute{\rm o}$rdova 3107, Vitacura, Santiago, Chile
   \and
   Department of Earth and Space Sciences, Chalmers University of Technology, Onsala Observatory, 439 94 Onsala, Sweden
   \and
   Argelander-Institut f\"{u}r Astronomie, Universit\"{a}t Bonn, Auf dem H\"{u}gel 71, D53121 Bonn
   \and
   National Radio Astronomy Observatory, 520 Edgemont Rd., Charlottesville, VA, 22903, USA
}
   \date{}

\authorrunning{Y. Ao et al.}
\titlerunning{The thermal state of molecular clouds in the Galactic Center}


\abstract{
We used the Atacama Pathfinder Experiment (APEX) 12 m telescope
to observe the $J_{\rm K_aK_c}$ = 3$_{03}\rightarrow2_{02}$,
3$_{22}\rightarrow2_{21}$, and 3$_{21}\rightarrow2_{20}$ transitions of
para-H$_2$CO at 218\,GHz simultaneously to determine kinetic temperatures of the dense
gas in the central molecular zone (CMZ) of our Galaxy. The map extends over
approximately 40${^\prime}$$\times$8${^\prime}$ ($\sim$100$\times$20\,pc$^2$)
along the Galactic plane with a linear resolution of 1.2\,pc. The strongest 
of the three lines, the H$_2$CO (3$_{03}\rightarrow2_{02}$) transition, is 
found to be widespread, and its emission shows a spatial distribution similar 
to ammonia. The relative abundance of para-H$_2$CO is 0.5$-$1.2$\times$10$^{-9}$, 
which is consistent with results from lower frequency H$_2$CO absorption lines. 
Derived gas kinetic temperatures for individual molecular clouds range from 50\,K 
to values in excess of 100\,K. While a systematic trend toward (decreasing) kinetic 
temperature versus (increasing) angular distance from the Galactic center (GC) is not 
found, the clouds with highest temperature ($T_{\rm kin}$ $>$ 100\,K) are all 
located near the nucleus. For the molecular gas outside the dense clouds, 
the average kinetic temperature is 65$\pm$10\,K. The high temperatures of molecular 
clouds on large scales in the GC region may be driven by turbulent energy
dissipation and/or cosmic-rays instead of photons. Such a non-photon-driven
thermal state of the molecular gas provides an excellent template for the more
distant vigorous starbursts found in ultraluminous infrared galaxies
(ULIRGs).

}
\maketitle


\section{Introduction}\label{introduction}

The Galactic center (GC) region is the closest galaxy core. It is
characterized by a high concentration of molecular gas located in the
innermost few hundred parsec of the Milky Way, the central molecular zone (CMZ)
(Morris \& Serabyn 1996), and by extreme conditions like high mass densities,
large velocity dispersions, strong tidal forces, and strong magnetic fields.
Therefore it is a unique laboratory for studying molecular gas in an environment
that is quite different from that of the Milky Way's disk. For a general
understanding of the physics involved in galactic cores, measurements of basic
physical parameters, such as molecular gas density and gas kinetic temperature,
are indispensable.

In local dark clouds, gas temperatures can be constrained by observations of
the $J$\,=\,$1\rightarrow0$ transition of CO, both because this transition is
opaque and easily thermalized and because the emission even fills the beam of a
single-dish telescope. At a distance of 8~kpc (Reid 1993), however, the beam
filling factor of CO $J$\,=\,$1\rightarrow0$ emission is unknown, and the GC
clouds may also be affected by self-absorption. Analysis of multilevel
studies of commonly observed linear molecules like CO, HCN, or HCO$^+$ suffers
from a coupled sensitivity to the kinetic temperature and gas density, making an
observed line ratio consistent with both a high density at a low
temperature and a low density at a high temperature. 
The metastable inversion lines of the symmetric top molecule ammonia (NH$_3$)
are frequently used as a galactic standard cloud thermometer (Walmsley \&
Ungerechts 1983; Danby et al. 1988).  Radiative transitions between K-ladders
of NH$_3$ are forbidden, and therefore the relative populations\footnote{NH$_3$
energy levels have quantum numbers ($J$,$K$), where $J$ denotes the total
angular momentum and K its projection on the molecule's axis. For metastable
levels, $J$\,=\,$K$.} depend on the kinetic temperature of the molecular gas
rather than its density. However, the fractional abundance of NH$_3$ varies
between 10$^{-5}$ in hot cores and 10$^{-8}$ in dark clouds (e.g. Benson \&
Myers 1983; Mauersberger et al. 1987).  Furthermore, NH$_3$ is extremely
affected by a high UV flux and tends to show a characteristic ``concave'' shape
in rotation diagrams, either caused by a variety of layers with different
temperatures (e.g., due to shocks) or by the specifics of the collision
rates.  Symmetric top molecules such as CH$_3$C$_2$H and CH$_3$CN are not
widespread and their emission is very faint (Bally et al. 1987; Nummelin et al.
1998).  Therefore we should look for a {\it widespread} symmetric or slightly
asymmetric top molecule that is more favorable for spectroscopic studies to
derive the kinetic temperature of the entire molecular gas. 

Formaldehyde (H$_2$CO) is such a molecule. It is truly ubiquitous. Wootten
\etal\,(1978) suggested that fractional H$_2$CO abundances decrease with
increasing density of the gas in star forming regions of the Galactic disk.
However, this may be due instead to decreasing volume filling factors with
increasing density (e.g., Mundy \etal\,1987). Unlike for NH$_3$,
variations in the fractional abundance of H$_2$CO rarely exceed one order of
magnitude (Johnstone et al. 2003). To give an example: the H$_2$CO abundance is
the same in the hot core and in the compact ridge of the Orion nebula, whereas
the NH$_3$ hot core abundance surpasses that of the ridge by about two orders
of magnitude (Caselli et al. 1993; Mangum et al. 1993). We also note that for
the starburst galaxy M\,82, $T_{\rm kin}$(NH$_3$) $\sim$ 60\,K, while for the
bulk of the molecular gas $T_{\rm kin}$(H$_2$CO) $\sim$ 200\,K (Wei\ss\, et al.
2001; Mauersberger et al. 2003; M{\"u}hle et al. 2007).

\begin{figure}[t]
\vspace{-0.0cm}
\centering
\includegraphics[angle=0,width=1.0\textwidth]{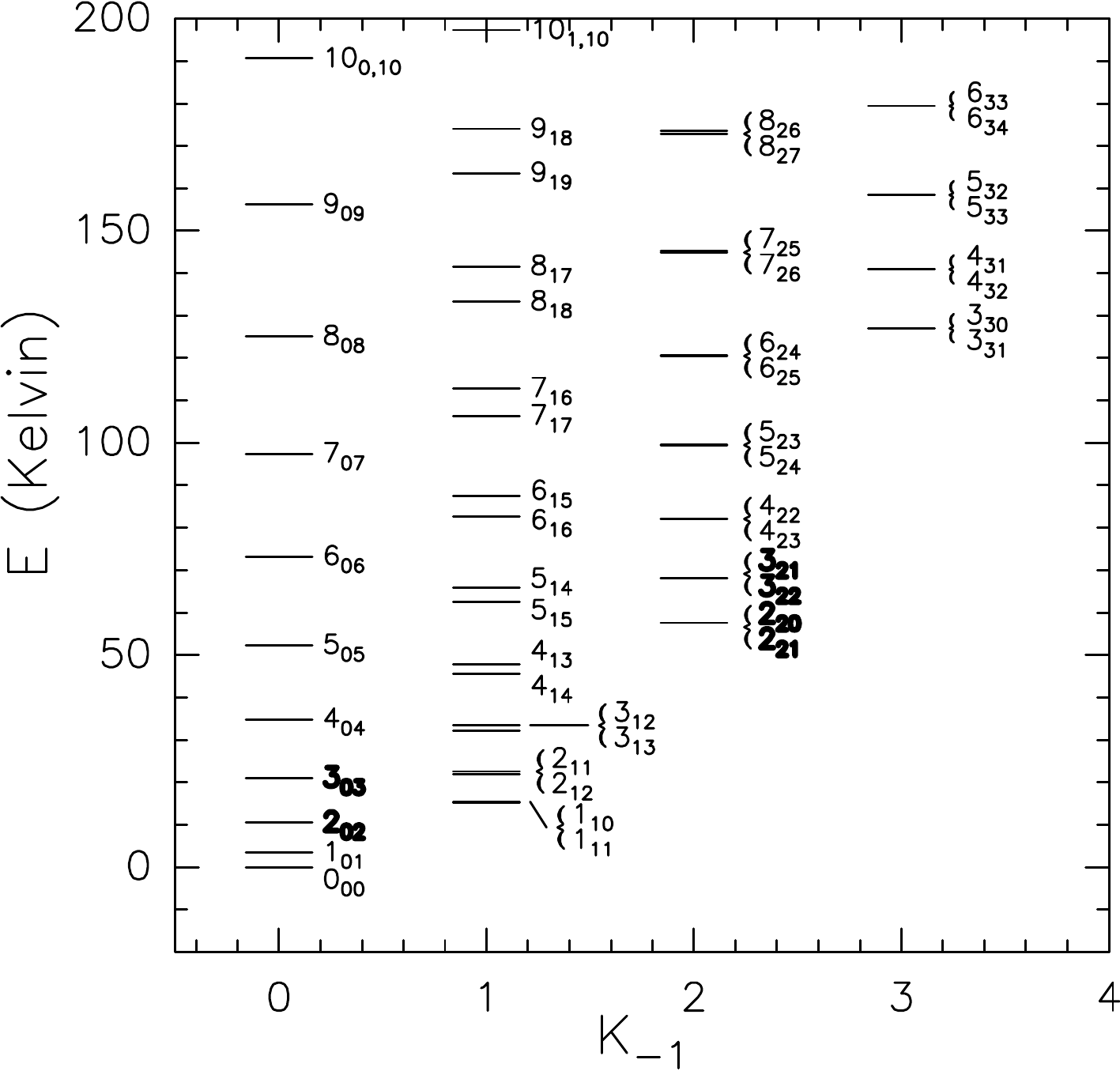}
\vspace{-0.0cm}
\caption{H$_2$CO energy-level diagram up to 200\,K. The H$_2$CO 218\,GHz transitions
observed in this paper are shown in bold.}
\label{h2co_level} 
\end{figure}

The relative populations of the $K_{\rm a}$ ladders of H$_2$CO (see Fig.~1
for an energy level diagram) are almost exclusively determined by collisional 
processes (Mangum \& Wootten 1993). Therefore, line ratios involving different 
$K_{\rm a}$ ladders of one of the subspecies, either ortho- or para-H$_2$CO, 
are good tracers of the kinetic temperature (Mangum \& Wootten 1993; M{\"u}hle 
et al. 2007). The energy levels above the ground (para)-H$_2$CO state are 
10.5 and 21.0\,K for the lower and upper states of H$_2$CO (3$_{03}\rightarrow2_{02}$),
and 57.6 and 68.1\,K for H$_2$CO (3$_{21}\rightarrow2_{20}$) and
(3$_{22}\rightarrow2_{21}$) (see Fig.~\ref{h2co_level} for all H$_2$CO energy levels
under 200\,K). Therefore, the line ratios are sensitive to gas
kinetic temperatures less than 100\,K, and the uncertainty in gas temperature
is relatively small for a measured line ratio at $T_{\rm kin}$\,$<$\,100\,K
(also see Mangum \& Wootten 1993) in the case of optically thin H$_2$CO
emission. At higher temperatures, the H$_2$CO 218\,GHz transitions are less
ideal because small changes in the ratios yield significant changes in $T_{\rm
kin}$, so that the H$_2$CO $J$\,=\,5$-$4 transitions at $\sim$364\,GHz are then
becoming better tracers (Mangum \& Wootten 1993).
The $J_{\rm K_aK_c}$ = 3$_{03}\rightarrow2_{02}$, 3$_{22}\rightarrow2_{21}$,
and 3$_{21}\rightarrow2_{20}$ transitions of para-H$_2$CO stand out by being
close in frequency. With rest frequencies of 218.222, 218.475, and
218.760\,GHz, respectively, all three lines can be measured simultaneously
by employing a bandwidth of 1\,GHz. In this way, the inter-ladder line ratios
H$_2$CO 3$_{22}\rightarrow2_{21}$/3$_{03}\rightarrow2_{02}$ and
3$_{21}\rightarrow2_{20}$/3$_{03}\rightarrow2_{02}$, are free of uncertainties
related to pointing accuracy, calibration errors, and different beam widths.
In this paper, we therefore present observations of the H$_2$CO line triplet at
218\,GHz to study the gas kinetic temperatures of Galactic center clouds.

\section{Observations and data reduction}\label{observation}

Simultaneous measurements of the $J_{\rm K_aK_c}$ = 3$_{03}\rightarrow2_{02}$,
3$_{22}\rightarrow2_{21}$, and 3$_{21}\rightarrow2_{20}$ transitions of
para-H$_2$CO (see $\S$\ref{introduction} and Fig.~1) were obtained 
with the Atacama Pathfinder Experiment (APEX\footnote{This publication is 
based on data acquired with the Atacama Pathﬁnder Experiment (APEX). APEX 
is a collaboration between the Max-Planck-Institut f{\"u}r Radioastronomie, 
the European Southern Observatory, and the Onsala Space Observatory.}, 
G{\"u}sten et al. 2006) 12 m telescope located on Chajnantor (Chile) between 
2010 April and 2010 June. We used the APEX-1 receiver, operating at 211$-$270 
GHz, which employs a superconductor-insulator-superconductor (SIS) mixer with 
a typical sideband rejection $>$\,10\,dB. As backends we used a fast fourier 
transform Spectrometer (FFTS, Klein et al. 2006), which consists of two units 
with a bandwidth of 1 GHz each and a channel separation of 244 kHz. The
full-width-at-half-maximum (FWHM) beam size was approximately 30$\arcsec$ in
the observed frequency range and the typical pointing error was $\sim$
3$\arcsec$. The main beam efficiency and the forward efficiency were 0.75 and
0.97, respectively.

We used the on-the-fly observing mode measuring
4${^\prime}$$\times$4${^\prime}$ maps in steps of 9$^{\prime\prime}$ in both
right ascension and declination, 0.8\,second integration time per position, and
one OFF position measurement after every two map rows, i.e., after
about one minute of observing time. The surveyed area is 302 square arcmin and
its dimension is roughly 40${^\prime}$$\times$8${^\prime}$ along the Galactic
plane. The total observing time was about 41 hours.

The data were reduced with the CLASS
software\footnote{http://www.iram.fr/IRAMFR/GILDAS}. We first excluded data
with a high noise level due to distorted baselines. The spectra were resampled
in steps of 15$\arcsec$ and smoothed to a velocity resolution of 2~\kms.  The
final maps comprise 4825 points, corresponding to 4825 spectra for each
transition. To optimize signal-to-noise ratios (SNRs) of integrated
intensity and channel maps, we had to determine the valid velocity ranges for
the spectra, especially for those cases where the lines were weak. To avoid
noise from channels without significant emission, we first created a high SNR
spectrum by smoothing all spectra within a 60$\arcsec$$\times$60$\arcsec$ box
to obtain a masking spectrum, for which 1 and 0 were assigned for the channels
with SNRs higher and lower than 5, respectively. Different line windows were
automatically deteremined to cover the emission from different positions, and the
baseline was removed before creating the masking spectrum.  Then, the final
spectra were created by multiplying the raw spectrum by the masking spectrum.
Because all lines were observed simultaneously and because the H$_2$CO
(3$_{03}\rightarrow2_{02}$) data have the best SNR, the masking spectrum from
this data set has also been applied to the other transitions by assuming the
same velocity ranges for the emission from these transitions (for the details
of this technique, see Dame et al. 2001; Dame 2011).

\section{Results}\label{results}

\subsection{General characteristics of the molecular gas}

\begin{figure}[t]
\vspace{-0.0cm}
\centering
\includegraphics[angle=0,width=1.0\textwidth]{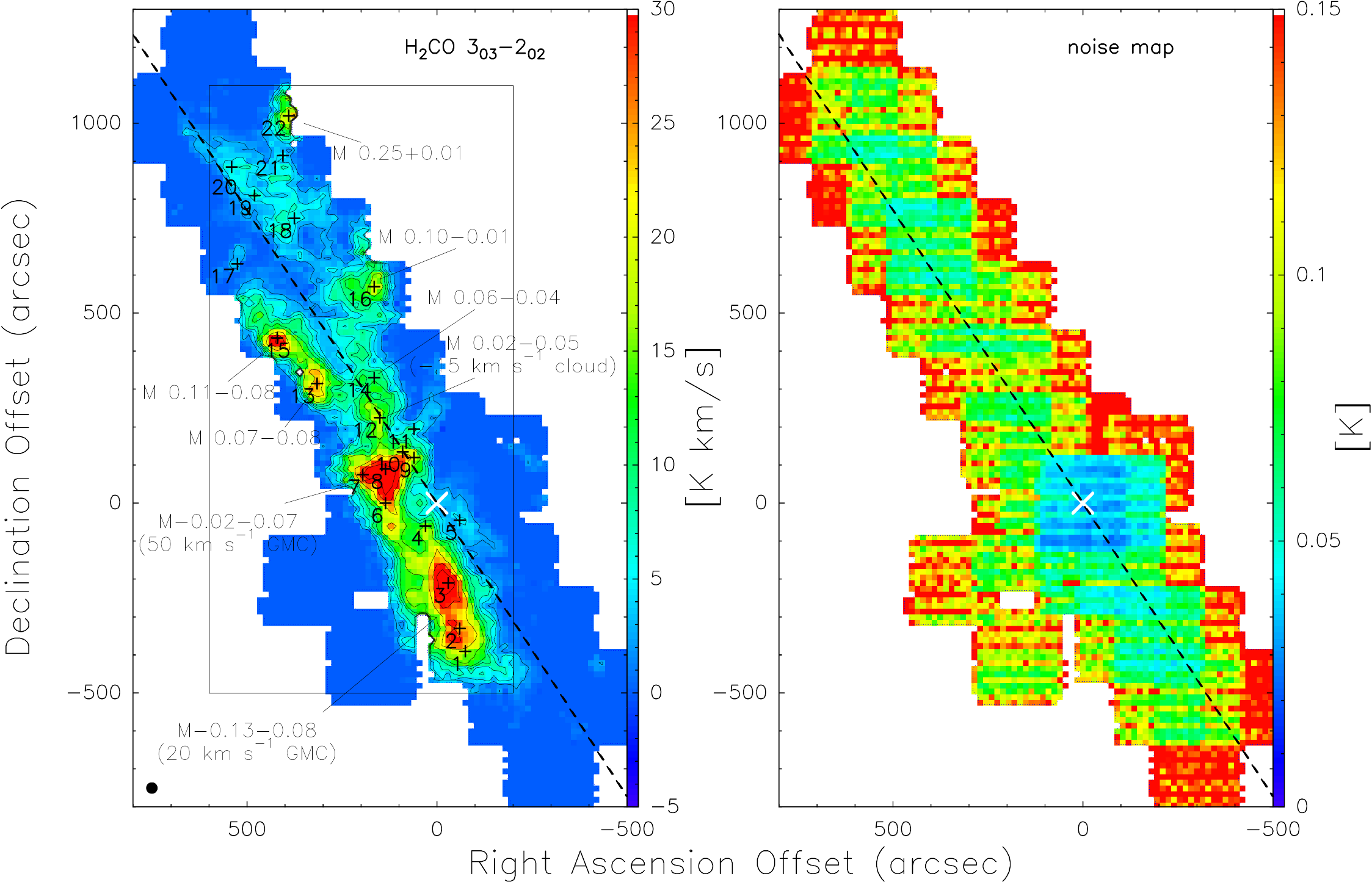}
\vspace{-0.0cm}
\caption{H$_2$CO (3$_{03}\rightarrow2_{02}$) integrated intensity map (left)
and the noise map (right) observed with the APEX in the GC. {\it Left:} black
contour levels for the molecular line emission (on a $T_{\rm A}^*$ scale) are
$-$3, 3, 6, 9, 12, 15, 20, 30, 40, 50~$\sigma$ (1$\sigma$\,=\,0.72~K~\kms). The
black rectangle shows a smaller region where most of the molecular line
emission is detected.  The molecular cloud labels with M followed by the
galactic coordinates are adopted from G{\"u}sten \etal\, (1981). The wedge at
the side shows the intensity range of the line emission on a $T_{\rm A}^*$
scale.  The beam size of 30$\arcsec$ is shown in the bottom-left corner.  {\it
Right:} Noise map across the region. The wedge at the side shows the range of
noise on a $T_{\rm A}^*$ scale at a velocity resolution of 2\,\kms. The spectra
presented in Fig.~\ref{spectra} are shown as the plus symbols with numbers.
Sgr\,A$^*$ is the reference position and shown as a white cross symbol. 
The dashed line marks the Galactic plane through Sgr\,A$^*$.}
\label{h2co303} 
\end{figure}

\begin{figure}[t]
\vspace{-0.0cm}
\centering
\includegraphics[angle=0,width=1.0\textwidth]{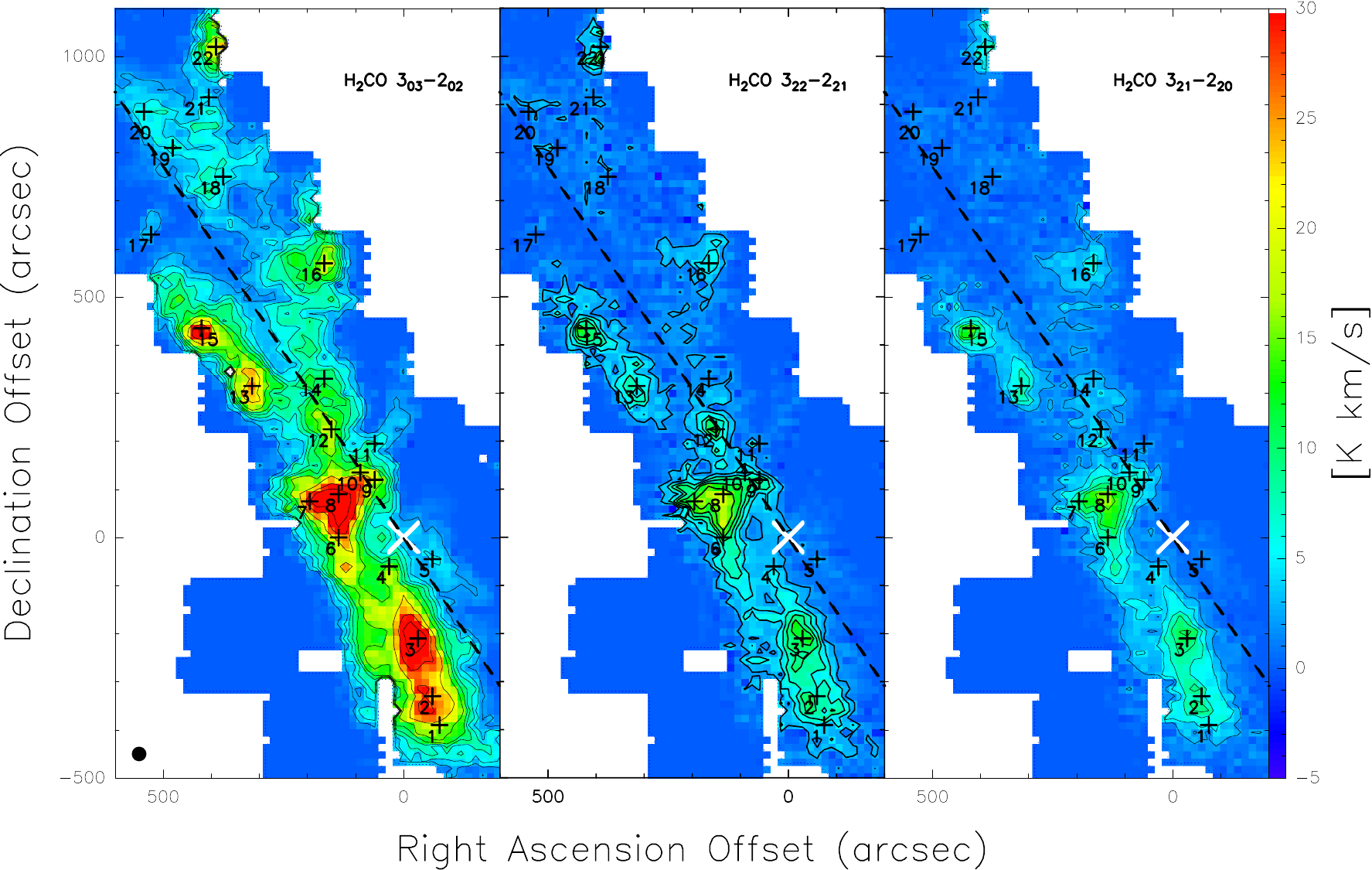}
\vspace{-0.0cm}
\caption{Integrated intensity maps for the different transitions observed in
the GC. The shown region corresponds to the rectangle (black solid line) in
Fig.~\ref{h2co303}~(left).  Black contour levels for the molecular line
emission (on a $T_{\rm A}^*$ scale) are $-$3, 3, 6, 9, 12, 15, 20, 30, 40,
50~$\sigma$ (1$\sigma$\,=\,0.72~K~\kms). The wedge at the side shows the intensity
range of the line emission on a $T_{\rm A}^*$ scale. The beam size
is shown at the bottom-left corner of the left panel.
The dashed line marks the Galactic plane through Sgr\,A$^*$.}
\label{h2co_all} 
\end{figure}

For the first time, observations of the triple transitions of H$_2$CO at
218\,GHz have been performed in a large area of the GC. Fig~\ref{h2co303} (left
panel) shows the extended line emission from the H$_2$CO
(3$_{03}\rightarrow2_{02}$) transition.  Molecular gas, revealed by H$_2$CO
(3$_{03}\rightarrow2_{02}$), shows a similar spatial distribution as ammonia
(G{\"u}sten et al. 1981). All prominent features identified in ammonia, e.g.,
the clouds M-0.13$-$0.08 and M-0.02$-$0.07 (the molecular clouds labeled with M
followed by the galactic coordinates are identified by G{\"u}sten \etal\,1981),
are clearly detected in H$_2$CO (3$_{03}\rightarrow2_{02}$), and labeled in the
figure.  A map providing the noise level of the full region is also presented
in Fig.~\ref{h2co303}~(right panel).  The median noise value for the 4825
spectra is about 0.1~K ($T_A^*$) at a velocity resolution of 2~\kms. In the
regions of interest where the line emission is strong, the noise level is
$\sim$ 0.08~K ($T_A^*$).  To show the line emission in more detail, we also
present integrated intensity maps (Fig.~\ref{h2co_all}), as well as channel
maps (see $\S$~\ref{data_cube}), for all transitions in a zoomed region in
Fig.~\ref{h2co303}~(left panel). The distribution of the velocity integrated
line emission of the observed three transitions is quite similar except that
the weaker two H$_2$CO transitions are not detected in some of the regions
where the 3$_{03}\rightarrow2_{02}$ transition is still observed.

\subsection{Individual spectral lines}\label{result_spec}
Spectra from some positions of interest are shown in Fig.~\ref{spectra} with
overlaid Gaussian fit profiles, and their locations are marked in
Fig.~\ref{h2co303}~(left panel). To achieve better SNRs, we averaged the
spectra within a 30$\arcsec$$\times$30$\arcsec$ area to create a new spectrum
for each transition. All transitions from a given position are presented in the
same panel of Fig.~\ref{spectra} but with different offsets along the y-axis.
For those positions with clear detections of H$_2$CO
3$_{22}\rightarrow2_{21}$ and 3$_{21}\rightarrow2_{20}$, line
parameters are listed in Table~\ref{table1}, where integrated intensity, $\int
T_{\rm mb} dv$, peak main beam brightness temperature, $T_{\rm mb}$, local
standard of rest (LSR) velocity, $V$, and FWHM
line width, $\Delta V_{1/2}$, were obtained from Gaussian fits.  In most cases,
only one component is needed for the Gaussian fits except in the case of
H$_2$CO (3$_{22}\rightarrow2_{21}$). This line is displaced from the
CH$_3$OH(4$_{22}\rightarrow3_{12}$) transition by only 49~\kms, and we
therefore have applied in this case two component Gaussian fits. 
 
In the central nuclear region, line emission is too weak to be detected with
sufficient signal-to-noise ratio, while H$_2$CO (3$_{03}\rightarrow2_{02}$) 
emission is detected southwest, $\sim$3\,pc away from \sgra\,, showing the broad 
weak line profile belonging to P5 ($-$60$\arcsec$, $-$45$\arcsec$) in 
Fig.~\ref{spectra}.

The line parameters obtained from Gaussian fits do not show many peculiarities
when one inspects the central velocities and line widths in Table~\ref{table1},
as well as the line profiles in Fig.~\ref{spectra}. Sometimes, however, there
are two velocity components, e.g., at offset positions P9 (60$\arcsec$,
120$\arcsec$) and P12 (150$\arcsec$, 225$\arcsec$) with respect to the Galactic
center, and the number of components for their Gaussian fits had to be doubled.
In particular, for the 44\kms\,component of P12, the H$_2$CO
(3$_{22}\rightarrow2_{21}$) integrated intensity seems to be significantly
larger than that of H$_2$CO (3$_{21}\rightarrow2_{20}$), because the H$_2$CO
(3$_{22}\rightarrow2_{21}$) transition is blended by the emission from
CH$_3$OH(4$_{22}\rightarrow3_{12}$) at 218.440 GHz. 

\subsection{Molecular line data cube}\label{data_cube}

\begin{figure}[t]
\vspace{-0.0cm}
\centering
\includegraphics[angle=0,width=1.0\textwidth]{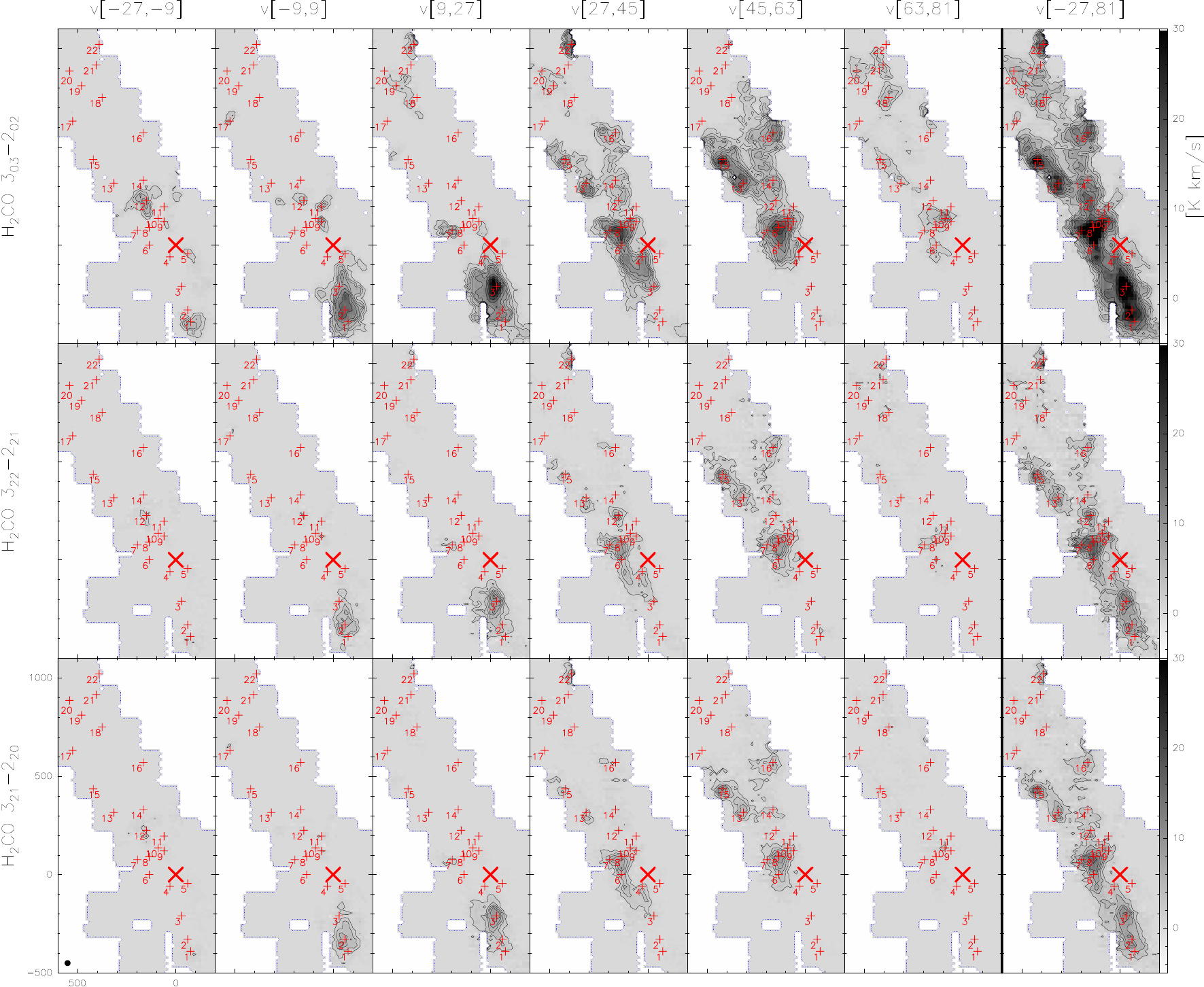}
\vspace{-0.0cm}
\caption{Selected velocity-integrated maps for the different transitions
observed in the GC. Black contour levels for the molecular line emission (on a
$T_{\rm A}^*$ scale) are $-$3, 3, 6, 9, 12, 15, 20, 30, 40, 50, 70~$\sigma$
(1$\sigma$\,=\,0.48~K~\kms) for the first six columns, and -3, 3, 6, 9, 12, 15,
20, 30, 40, 50~$\sigma$ (1$\sigma$\,=\,0.72~K~\kms) for the last column,
respectively.  The wedges at the sides show the intensity scale of the line
emission on a $T_{\rm A}^*$ scale.}
\label{h2co_selected} 
\end{figure}

In Fig.~\ref{h2co_selected}, channel maps with velocity steps of 18~\kms\, are 
presented, allowing us to clearly separate the components based on both
their positions and velocities. The total integrated intensity maps are also
shown for comparison in the panels on the righthand side. Channel maps in steps
of 2~\kms\, are presented in the Appendix
(Fig.~\ref{h2co303_channel}-\ref{h2co321_channel}) to show more detail.

We choose the H$_2$CO (3$_{03}\rightarrow2_{02}$) data to describe the
individual molecular concentrations.  H$_2$CO 3$_{03}\rightarrow2_{02}$
emission mainly ranges from $-$27 to 81~\kms.  In the velocity range
[$-$27,$-$9], there are two prominent features, the $-$15~\kms\, cloud
M~0.02$-$0.05 at P12, and the southernmost cloud at P1.
The latter is part of the 20\kms\, cloud M-0.13$-$0.08. 
There is also weak H$_2$CO 3$_{03}\rightarrow2_{02}$ emission at
P5 close to \sgra, which traces the southwestern lobe of the circumnuclear disk
(CND).

Within the velocity range [$-$9,9], the bulk of the 20\kms\, cloud appears in
the south and its size is about 7\,pc$\times$15\,pc. There are three {\bf small
cores} at P9, P11, and P12 with sizes of $\sim$ 1 to 2\,pc in the
northeast and one small core at P17 even farther away from \sgra. The peak 
of the 20\kms\, cloud moves from south ($-$75$''$, $-$390$''$) relative to the 
Galactic center to north ($-$30$''$, $-$210$''$), increasing the velocity from 
$-$27 to $+$27~\kms. 

Within the velocity range [9,27], there is a dense concentration at P7
with a size of $\sim$ 2.7\,pc$\times$5.4\,pc. Extended weak emission is detected
around P19 in an irregular morphology with a size of $\sim$ 2\,pc$\times$9\,pc.
The northernmost core, M~0.25+0.01, begins to appear at P22. This
concentration is not fully covered by our observations.

In the velocity range [27,63], the line emission shows a complex morphology.
The prominent features are the 50\kms\, cloud M-0.02$-$0.07 around P6, P7, and
P8, two compact concentrations M~0.07$-$0.08 at P13 and M~0.11$-$0.08 at P15,
an extended region of weak emission associated with M~0.06$-$0.04 around P12
and P14, a concentration M~0.10$-$0.01 around P16, and the northernmost core
M~0.25+0.01 at P22. In the velocity range [27,45], the gas in the southeast
close to \sgra\, appears to connect the 20~\kms\, and 50~\kms\, clouds.

In the extreme velocity range [63,81], the 50 \kms\, cloud moves to the west by
$\sim$60$''$ with respect to [27,63], and peaks at P10, the edge of the cloud.
In addition to features identified in the previous velocity ranges, there are
three clumps around P18, P20, and P21, with sizes of about 3\,pc$\times$6\,pc.

\section{Discussion}

\subsection{Formaldehyde column density and abundance}\label{result_abundance}
In the following we derive H$_2$CO column densities and abundances. Assuming
the line emission is optically thin and the contribution from the cosmic
microwave background is negligible, the H$_2$CO column density, $N(J_{\rm
K_aK_c})$, in an upper state $J_{\rm K_aK_c}$ can be obtained by 
\begin{equation}
N(J_{\rm K_aK_c})\,=\,\frac{3k}{8\pi^3\nu S\mu^2}\,\frac{h\nu/kT_{\rm
ex}}{e^{h\nu/kT_{\rm ex}}-1}\,\int T_{\rm mb}\,d{\rm v}\,,
\end{equation}
(for the equations in this subsection, see Mangum \& Shirley
2008\footnote{https://safe.nrao.edu/wiki/pub/Main/MolInfo/column-density-calculation.pdf}),
where $h$ is Planck's constant, $k$ denotes Boltzmann's constant, $\mu$ is the
dipole moment, $\nu$ the frequency of the transition, $T_{\rm ex}$ the
excitation temperature, $S$ the line strength, $T_{\rm mb}$ the main beam
brightness temperature, and $\int T_{\rm mb}\,d{\rm v}$ the integrated line
intensity for the transition $J_{\rm K_aK_c}\rightarrow(J-1)_{\rm K_aK_c-1}$.
H$_2$CO is a slightly asymmetric top molecule, and its line strength, $S$, can
be approximately calculated as for a symmetric top molecule by
\begin{equation}
S\,=\,\frac{J^2-K_a^2}{J(2J+1)}\,.
\end{equation}
The total column density, $N_{\rm total}$, is related to the column density,
$N(J_{\rm K_aK_c})$, in the upper state $J_{\rm K_aK_c}$ by
\begin{equation}
\frac{N_{\rm total}}{N(J_{\rm K_aK_c})}\,=\,\frac{Z}{g{\rm _J}g{\rm _K}g{\rm
_I}}\,{\rm exp}({\frac{E(J_{\rm K_aK_c})}{kT_{\rm ex}}})\,,
\end{equation}
where $g_{\rm J}$ (=2J+1) is the rotational degeneracy, $g_{\rm K}$ marks the $K$
degeneracy, $g_{\rm I}$ the nuclear spin degeneracy, $E(J_{\rm K_aK_c})$ the
energy of state $J_{\rm K_aK_c}$ above the ground level, and $Z$ the partition 
function. The partition function $Z$ can be calculated by
\begin{equation}
Z\,=\,\sum^{\infty}_{J=0}\,\sum^{J}_{K_a=0}\,\sum^{J-K_a+1, K_c\leq J}_{K_c=J-K_a}\,g{\rm _J}g{\rm _K}g{\rm _I}\,{\rm
exp}({-\frac{E(J_{\rm K_aK_c})}{kT_{\rm ex}}})\,.
\end{equation}
For para-H$_2$CO, $g_{\rm K}$=1, $g_{\rm I}$=1, and $K_{\rm a}$ can only be
even. Here we include 41 levels with energies above ground state up to 286\,K
and assume the same excitation temperatures for all transitions (i.e., 
local thermodynamical equilibrium, LTE) to estimate the partition function. 
Substituting units and parameters for H$_2$CO (3$_{03}\rightarrow2_{02}$)\, 
in Equation\,(3), and assuming an excitation temperature of 10\,K, the total 
para-H$_2$CO column density, $N_{\rm total}$, is 
\begin{equation}
N_{\rm total}\,=\,1.32\times 10^{12}\int T_{\rm mb}(3_{03}\rightarrow2_{02})\,dv\,{\rm cm^{-2}}\,,
\end{equation}
where the integrated line intensity, $\int T_{\rm mb}(3_{03}\rightarrow2_{02})\,d\rm v$, is
in units of K\,\kms.

For the \sgra\,complex with a size of 30\,pc in diameter, corresponding to
$\sim$\,13\,arcmin, the estimated molecular gas mass is
$\sim$0.4$\times$10$^6$\,\msun\, (Kim \etal\,2002), yielding an average H$_2$
column density of 2.6$\times$ 10$^{22}$\,cm$^{-2}$.  In this region, 2161
formaldehyde spectra were obtained with a spacing of 15$\arcsec$, and the
average spectrum has an integrated intensity of 10.5\,K\,\kms\, on a $T_{\rm
mb}$ scale. The derived averaged para-H$_2$CO column density is
1.4$\times$10$^{13}$\,cm$^{-2}$ assuming an excitation temperature of 10\,K.
Adjusting the excitation temperature in the range of 5 to 40\,K, the resulting
total para-H$_2$CO column density will decrease with increasing $T_{\rm ex}$
from 5\,K to around 14\,K, because with increasing $T_{\rm ex}$ the populations
of the $J$ = 2 and 3 states become more dominant. Beyond $T_{\rm ex}$ = 14\,K
the resulting column densities will increase with excitation temperature
because then also the $J$ $>$ 3 levels will be populated.  For $T_{\rm kin}$
$\la$ 100\,K (see $\S~\ref{dis_lvg}$ for large velocity gradient (LVG)
modeling), $T_{\rm ex}$ values $\ga$40\,K require densities
$\ga$10$^6$\,cm$^{-3}$, which are unrealistically high on a large spatial scale
(e.g., G{\"u}sten \& Henkel 1983). Therefore higher $T_{\rm ex}$ values can
probably be excluded and the corresponding averaged para-H$_2$CO column density
can be constrained to (1.3$-$3.1) $\times$ 10$^{13}$\,cm$^{-2}$.  The resulting
para-H$_2$CO abundance is (0.5$-$1.2) $\times$ 10$^{-9}$. This abundance agrees
with the values found by G{\"u}sten \& Henkel (1983) and Zylka et al. (1992),
who used $K_{\rm a}$\,=\,1 ortho-formaldehyde K-doublet absorption lines to
obtain H$_2$CO abundances of $\sim$10$^{\rm -10}$ to 2$\times$10$^{\rm -9}$ in
the Galactic center region.  Thus it is reasonable to adopt two fixed limiting
para-H$_2$CO abundances of 10$^{\rm -9}$ and 10$^{\rm -10}$ for the LVG
analysis in $\S~\ref{dis_lvg}$.


\subsection{Line ratios}\label{result_ratio}

\begin{figure}[t]
\vspace{-0.0cm}
\centering
\includegraphics[angle=0,width=1.0\textwidth]{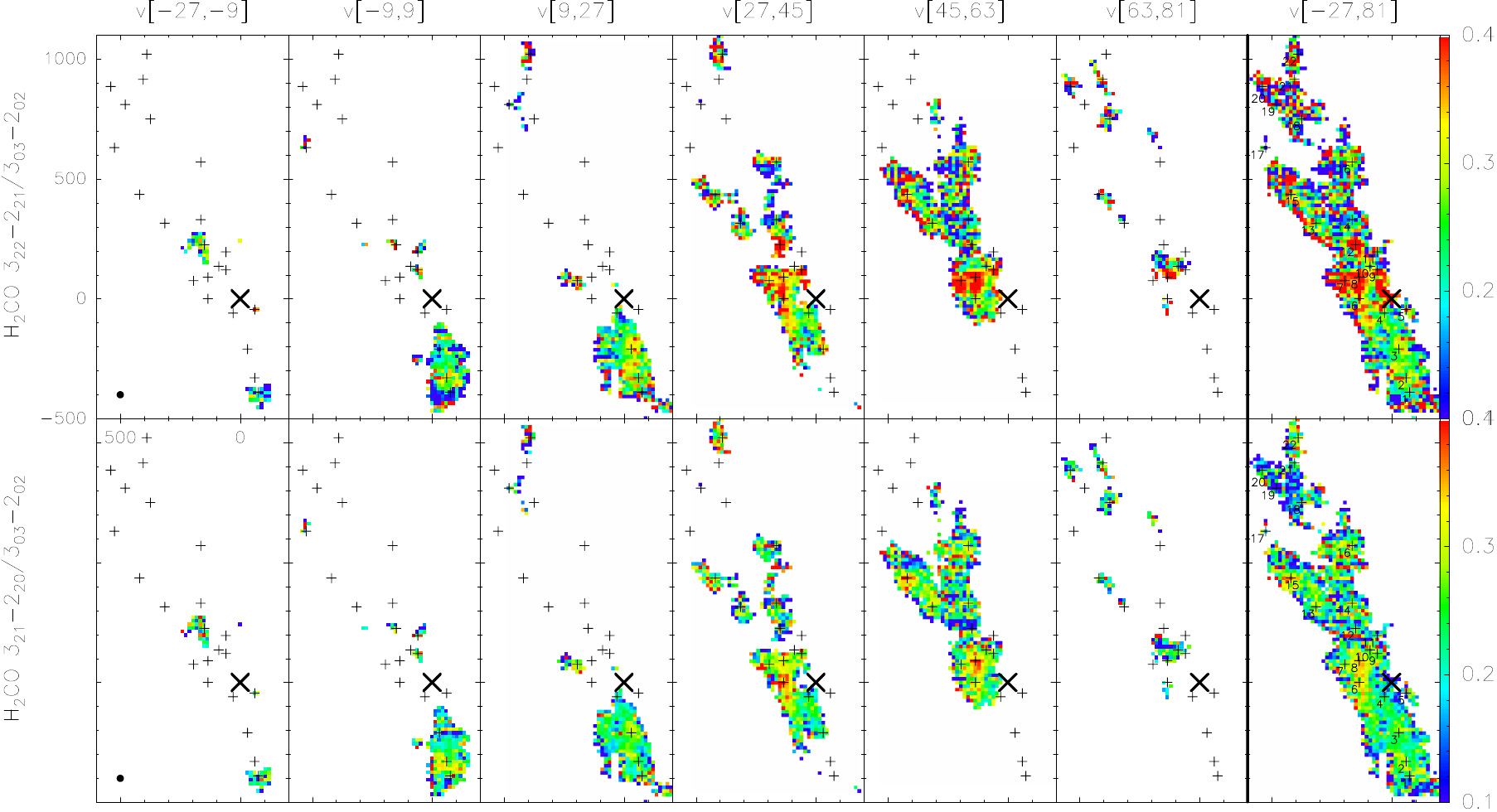}
\vspace{-0.0cm}
\caption{H$_2$CO 3$_{22}\rightarrow2_{21}$/3$_{03}\rightarrow2_{02}$ (top) and  
3$_{21}\rightarrow2_{20}$/3$_{03}\rightarrow2_{02}$ (bottom) integrated intensity ratio
map. Ratios are calculated when the H$_2$CO 3$_{03}\rightarrow2_{02}$ line emission is
detected above 5 $\sigma$. 
The top and bottom rows should be nearly identical, and the difference mainly
comes from the CH$_3$OH contamination in H$_2$CO 3$_{22}\rightarrow2_{21}$.
The wedges at the sides show the line ratios.
The beam size of 30$\arcsec$ is shown at the bottom-left corner. 
Sgr\,A$^*$ is the origin for the offset coordinates and shown as a cross.}
\label{ratio} 
\end{figure}

In this survey, the three 218\,GHz rotational transition lines of H$_2$CO are
observed simultaneously at the same angular resolution, providing good data
sets to derive the H$_2$CO line ratios. Before quantitatively determining gas
kinetic temperatures, we first present H$_2$CO
3$_{22}\rightarrow2_{21}$/3$_{03}\rightarrow2_{02}$ and
3$_{21}\rightarrow2_{20}$/3$_{03}\rightarrow2_{02}$ line ratios in
Fig.~\ref{ratio} as probes of gas temperature. The line ratio maps are derived
from both channel maps and total integrated intensity maps, and the ratios are
calculated by integrating channels where the H$_2$CO(3$_{03}\rightarrow2_{02}$)
line emission is detected above 5~$\sigma$. Since the excitation conditions
for 3$_{22}\rightarrow2_{21}$/3$_{03}\rightarrow2_{02}$ and
3$_{21}\rightarrow2_{20}$/3$_{03}\rightarrow2_{02}$ are very similar, these two
line ratio maps should be nearly identical, and indeed the differences mainly arise
from the CH$_3$OH contamination in H$_2$CO 3$_{22}\rightarrow2_{21}$.

As seen in Fig.~\ref{ratio}, the H$_2$CO
3$_{21}\rightarrow2_{20}$/3$_{03}\rightarrow2_{02}$ line ratio, with a median
value of 0.23, varies significantly across the mapped region, from about 0.15
at the edge of the clouds to $\sim$0.35 toward the 20\,\kms\, GMC and the compact
concentration M 0.11$-$0.08 at P15. In case of narrow one-component features, 
the 3$_{22}\rightarrow2_{21}$/3$_{03}\rightarrow2_{02}$ ratios follow the same
trend as the 3$_{21}\rightarrow2_{20}$/3$_{03}\rightarrow2_{02}$ ratios. 
For the 20 \kms\, GMC, the line ratio is higher in the south 
than in the north in the velocity range [$-$9,9]. Within the velocity range 
[9,27], the ratio becomes higher at P3 in the north of the 20 \kms\, GMC. 
The molecular clouds around P18, P19, and P20 have two velocity components
as presented in Fig.~\ref{spectra}, one at $\sim$\,25\,\kms\, and another at
$\sim$\,78\,\kms, and are characterized by low line ratios as clearly shown in
Fig.~\ref{ratio}\,(bottom). Higher line ratios tend to suggest higher gas
kinetic temperatures and vice versa because the relative populations of the
$K_{\rm a}$ ladders of H$_2$CO are almost exclusively determined by collisional
processes (Mangum \& Wootten 1993). To be more quantitative and to relate the
line ratios to kinetic temperatures, we need to adopt LVG radiative transfer
modeling, which is done in the following section.

\subsection{Kinetic temperatures of the Galactic center clouds}\label{dis_lvg}
To evaluate gas kinetic temperatures, we selected the positions with
the Gaussian fits listed in Table~\ref{table1} (see also Fig.~\ref{spectra}). 
To investigate the gas excitation from the H$_2$CO line measurements, we use 
a one-component LVG radiative transfer model with collision rates from Green (1991)
and choose a spherical cloud geometry with uniform kinetic temperature and density 
as described in Mangum \& Wootten (1993). 
Dahmen \etal\,(1998) estimated that the 
velocity gradient ranges from 3 to 6\,\kmspc\, for Galactic center clouds. Here 
we adopt two fixed para-H$_2$CO abundances of [para-H$_2$CO] = 10$^{\rm -9}$ and 
10$^{\rm -10}$ (see $\S$~\ref{result_abundance}), and a velocity gradient of 
$\rm{(dv/dr)}$\,=\,5 ($\rm \kms$)\,pc$^{-1}$. The modeled parameter space 
encompasses gas temperatures, $T_{\rm kin}$, from 10 to 300\,K with a step 
size of 5\,K and H$_2$ number densities per cm$^{3}$, log\,$n_{\rm H_2}$, 
from 3.0 to 7.0 with a logarithmic step size of 0.1. According to Green (1991), 
collisional excitation rates for a given transition are accurate to $\sim$20\%.

We first choose the H$_2$CO 3$_{22}\rightarrow2_{21}$ and
3$_{03}\rightarrow2_{02}$ data as input parameters. Although the
3$_{22}\rightarrow2_{21}$ transition is blended with CH$_3$OH at a few
positions, its line emission is stronger than that of H$_2$CO
3$_{21}\rightarrow2_{20}$, and the components can be separated in all cases.
This even holds for position P12 (see $\S$~\ref{result_spec}), where we have
determined the central velocities of the two velocity components from H$_2$CO
3$_{03}\rightarrow2_{02}$ to then carry out a four-component Gaussian fit to
both H$_2$CO 3$_{03}\rightarrow2_{02}$ and CH$_3$OH 4$_{22}\rightarrow3_{12}$.
Comparing computed line intensities and their line ratios with the
corresponding observational results, we can constrain the kinetic temperature.
The gas density is not well known because it is highly dependent on the adopted
fractional abundance, velocity gradient, and filling factors.  Here we choose a
filling factor of unity to fit the data, implying that we obtain beam averaged
quantities. 

\begin{figure}[t]
\vspace{-0.0cm}
\centering
\includegraphics[angle=0,width=1\textwidth]{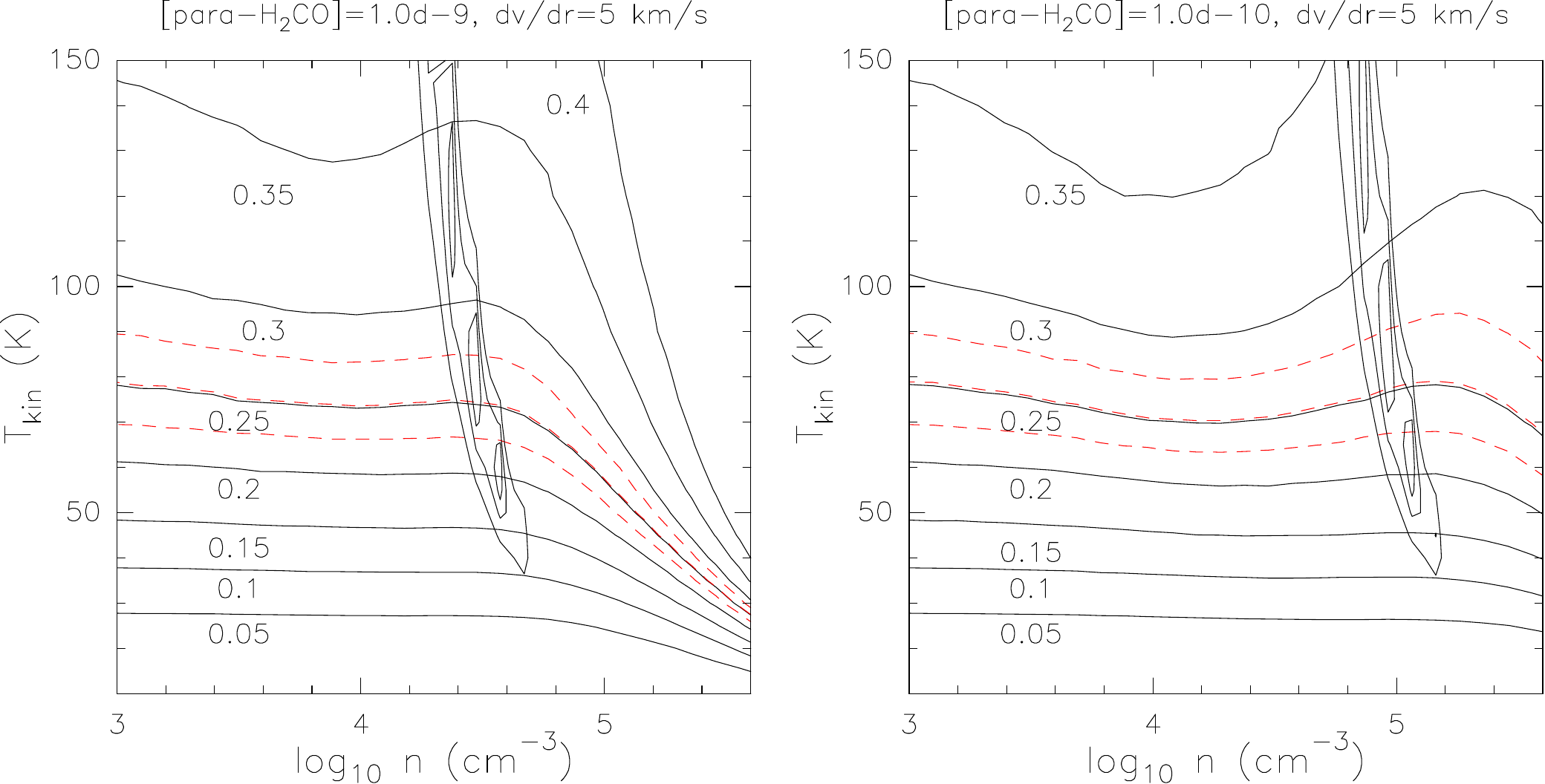}
\includegraphics[angle=0,width=1\textwidth]{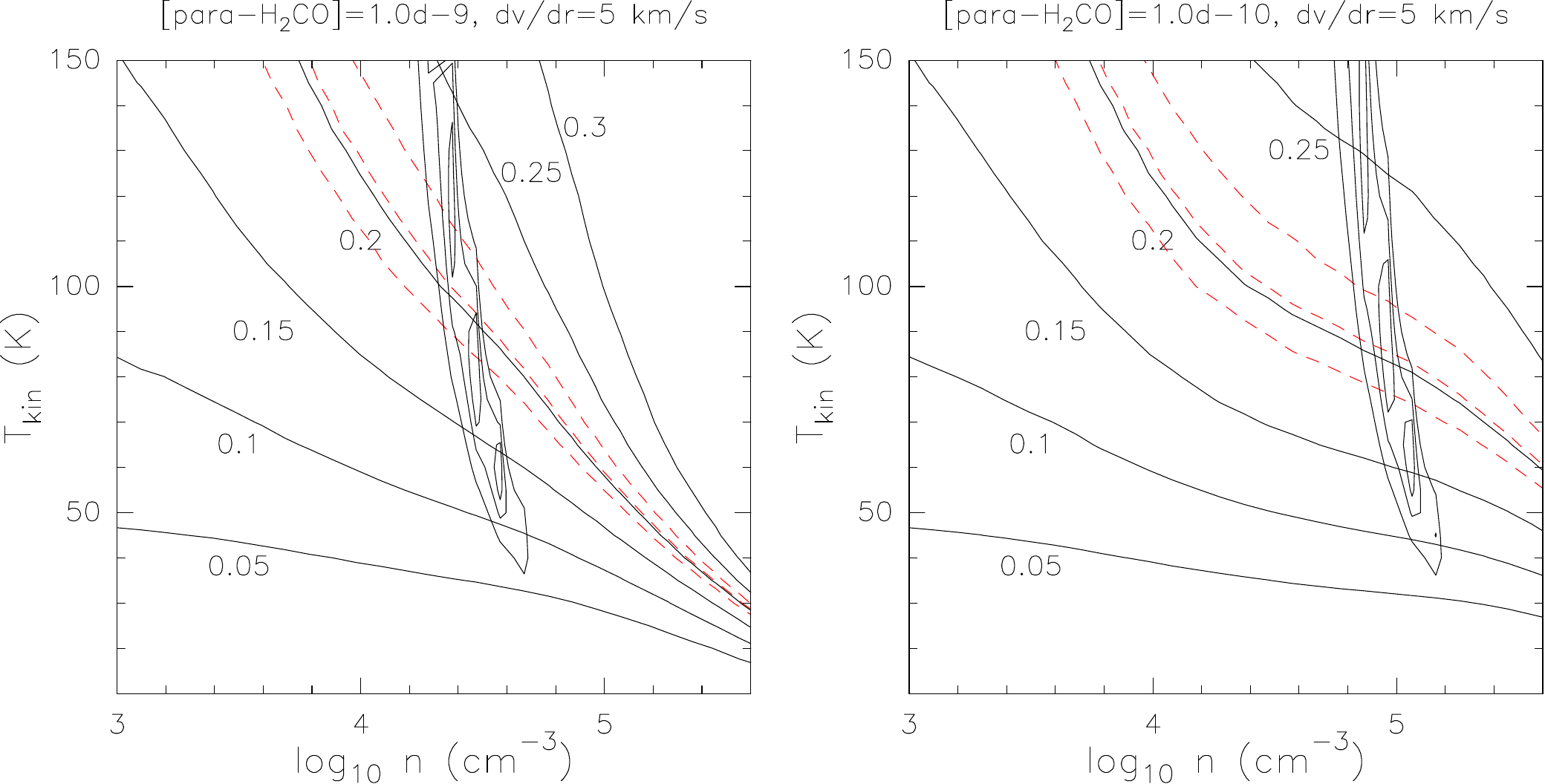}
\vspace{-0.0cm}
\caption{Example of LVG modeling for P4. {\it Top:} reduced
$\chi^2$ distribution (mainly vertical contours) for a single-component LVG
model fit to the H$_2$CO brightness temperatures (black contours,
$\chi^2$\,=\,1,2,4), as well as H$_2$CO
3$_{22}\rightarrow2_{21}$/3$_{03}\rightarrow2_{02}$ line ratios (mainly
horizonal contours) as a function of $n_{\rm H_2}$ and $T_{\rm kin}$. The solid
lines represent the line ratios: 0.05, 0.1, 0.15, 0.2, 0.25, 0.3, 0.35, 0.4.
The red dashed lines show the observed line ratio and its lower and upper
limits. The para-H$_2$CO abundances per velocity gradient,
[para-H$_2$CO]/$\rm{(dv/dr)}$, for the LVG models are 2$\times$10$^{\rm -10}$
pc\,($\rm \kms$)$^{-1}$ (left) and 2 $\times$ 10$^{-11}$\,pc\,($\rm
\kms$)$^{-1}$ (right), respectively. In the left panel the lines with
a given H$_2$CO line ratio move downwards (lower $T_{\rm kin}$) at high density
because the H$_2$CO lines start to become saturated; this causes intensity
ratios for a given $T_{\rm kin}$ to get closer to unity. 
{\it Bottom:} reduced $\chi^2$ distribution for the H$_2$CO
3$_{21}\rightarrow2_{20}$/3$_{03}\rightarrow2_{02}$ line ratios.
The kinetic temperature is sensitive to the gas density so this line
ratio is a less suitable thermometer.}
\label{lvg} 
\end{figure}

In Fig.~\ref{lvg}(top), an example is presented to show how the parameters are
constrained by the reduced $\chi^2$ distribution of H$_2$CO line measurements
in the $T_{\rm kin}$-$n$ parameter space. As can be seen, observed H$_2$CO
3$_{22}\rightarrow2_{21}$/3$_{03}\rightarrow2_{02}$ line ratios (the
approximately horizontal lines in the diagram) are a good measure of $T_{\rm
kin}$, independent of density, as long as the transitions are optically thin.
In contrast, the density is poorly constrained, as can be inferred by the
different resulting densities for the chosen limiting fractional H$_2$CO
abundances (in the specified case of Fig.~\ref{lvg}, the range covers a factor
of 3$-$4 if assuming a filling factor of unity). In Table~\ref{table2}, we
present derived gas kinetic temperatures averaged over 30$\arcsec$ boxes at the
positions shown in Figs.~\ref{h2co303} and \ref{spectra}. It is worth noting
that in most cases the two adopted para-H$_2$CO abundances, differing by a
factor of 10 ($\S$4.1), cause only a slight change of less than 10\,K in
kinetic temperature because the 218\,GHz H$_2$CO transitions remain optically
thin in most parameter ranges except for gas densities higher than
10$^{4.6}$\,cm$^{-3}$ in the case of [para-H$_2$CO] = 10$^{-9}$. For
[para-H$_2$CO] = 10$^{-10}$ this limiting density is well above 10$^{5}$ and
outside the plotted range of densities of Fig.~\ref{lvg}(top).  This
demonstrates that H$_2$CO is a good molecular thermometer and can be used to
reliably determine kinetic temperatures. 

Gas temperatures range from 50~K in the southern part of the 20\kms\, cloud to
above 100~K in the 50\kms\, cloud and the molecular core of M~0.07$-$0.08.
While a systematic trend of (decreasing) kinetic temperature versus
(increasing) angular distance from the nucleus is not found, the clouds with
highest temperature ($T_{\rm kin}$ $>$ 100\,K) are all located near the center.
To estimate the overall gas temperature on a large scale of $\sim$90\,pc, we
mask the dense clouds by clipping all the emission above 3\,$\sigma$ in $T_{\rm
A}^*$ of the H$_2$CO (3$_{03}\rightarrow2_{02}$) line within a
60$\arcsec$$\times$60$\arcsec$ box and create an averaged spectrum, yielding a
gas kinetic temperature of 65\ppm10\,K for more diffuse molecular gas outside
of the dense cores in the GC region.

If we instead use the H$_2$CO
3$_{21}\rightarrow2_{20}$/3$_{03}\rightarrow2_{02}$ line ratio to constrain gas
properties, the kinetic temperature is somewhat more sensitive to the gas
density so that this line ratio is not quite as good as a thermometer (see
Fig.~\ref{lvg}(bottom)). Therefore, we focus exclusively on the H$_2$CO
3$_{22}\rightarrow2_{21}$/3$_{03}\rightarrow2_{02}$ line ratio to derive the
gas kinetic temperature in this study.

High gas temperatures were first deduced from the metastable 
transitions of ammonia (G{\"u}sten \etal\,1981, 1985; H{\"u}ttemeister \etal\,1993). 
Using the CO 7$-$6/4$-$3 line ratio, Kim \etal\,(2002) report, for the Sgr\,A 
complex, a gas kinetic temperature of 47~K on a linear scale of 30\,pc. 
From mm- and submm-line spectroscopy, Oka \etal\,(2011) deduced temperatures 
of at least 63\,K in the CND. The 20 \kms\, and 50 \kms\, clouds were studied 
by G{\"u}sten \etal\,(1981, 1985) with species like NH$_3$ and CH$_3$CN
and by Mauersberger \etal\,(1986) in the ($J$,$K$)\,=\,(7,7) metastable inversion 
line of ammonia, yielding gas temperatures in the range 80$-$100~K. These results 
are roughly consistent with the temperatures derived by us from H$_2$CO.
We emphasize, however, that ammonia may be more affected than H$_2$CO by a
peculiar molecule specific chemistry and that the degeneracy between high
$T_{\rm kin}$ and low $n$(H$_2$) or vice versa is difficult to overcome for the
mm- and submm-transitions from linear molecules. 

A high gas temperature will affect the Jeans masses of dense cores (e.g., for a molecular
cloud, its Jeans' mass is
$M_{\rm J}$\,=\,$(\frac{5\,k\,T_{\rm kin}}{G\,\mu\,2m_p})^{1.5}\,(\frac{4\,\pi}{3})^{-0.5}\,\rho^{-0.5}$
\,=\,$1.25\,(\frac{T_{\rm kin}}{10\,{\rm K}})^{1.5}\,(\frac{n}{10^5\,{\rm cm^{-3}}})^{-0.5}$\,\msun,
where $n$ is volume density and $T_{\rm kin}$ is gas kinetic temperature. 
$M_{\rm J}$ $\sim$1.2\,\msun\,for $T_{\rm kin}$\,=\,10\,K and $n$\,=\,10$^5$\,cm$^{-3}$, and
14$-$39\,\msun\,for $T_{\rm kin}$\,=\,50$-$100\,K and $n$\,=\,10$^5$\,cm$^{-3}$),
and may affect the initial mass function (IMF) of star formation, resulting in
a top-heavy IMF in the Galactic center region (Klessen et al. 2007). Indeed,
Alexander \etal\,(2007) suggest a top-heavy IMF to explain the observed ring of
massive stars orbiting about 0.1 pc around the Galactic center.  However,
Bartko \etal\,(2011) do not find evidence of a top-heavy IMF at distances beyond
12$\arcsec$ from Sgr\,A$^*$.  Higher densities in the GC clouds with respect to
those in the spiral arms due to a higher stellar density and strong tidal
forces may lead to lower Jeans masses and may thus counteract the effect of
higher temperatures. 

\subsection{Heating mechanisms in the GC: turbulent heating or cosmic-ray heating}\label{dis_heating}

What heats the dense gas to high temperatures in the GC?  The four most 
common mechanisms for heating the gas in molecular clouds are (a) photo-electric 
heating in photon-dominated regions (PDRs), (b) X-ray heating (XDRs), 
(c) cosmic-ray heating (CRDRs), and (d) turbulent heating. In star-forming 
regions, gas can be heated by electrons released from normal dust grains or polycyclic 
aromatic hydrocarbons (PAHs).  Gas and dust are thermally coupled in very dense regions
($n_{\rm H_2}$$>$10$^5$\,cm$^{-3}$; e.g., Kr{\"u}gel \& Walmsley 1984).
However, the gas temperatures derived from H$_2$CO are much higher than the
fairly uniform dust temperatures of $\sim$ 14$-$20~K in the dense clouds of the GC
(Pierce-Price \etal\,2000; Garc{\'{\i}}a-Mar{\'{\i}}n et al. 2011; Molinari et
al. 2011). Photons can drive such decoupling only at the very surface of
irradiated clouds ($\sim$ a few percent of their total molecular gas mass; see
Bradford et al. 2003 and references therein). Only there might they have a
chance to dissociate complex molecules, such as H$_2$CO, which then necessarily
probe much deeper and UV-shielded gas regions where photo-electric heating of
the gas is no longer dominant.  Therefore, some other 
process(es) should exist that are efficient at directly heating the gas from outside.  X-ray
heating (Maloney \etal\,1996), cosmic-ray heating (G{\"u}sten \etal\,1981,
1985; Papadopoulos 2010), and turbulent heating (G{\"u}sten \etal\,1985; 
Schulz \etal\,2001; Pan \& Padoan 2009) are such potential heating mechanisms.

A few other heating processes also deserve to be mentioned. 
Gravitational heating (Goldsmith \& Langer 1978; Tielens 2005) is important 
during the collapse phase of molecular cloud cores. However, gravitational 
collapse is a temporary phase and star-forming activity, with the notable 
exception of Sgr\,B2 (not covered by our survey), is not vigorous in 
the clouds of the CMZ. Heating of the gas by magnetic ion-neutral slip 
(Scalo 1977; Goldsmith \& Langer 1978) would require further observations 
of the ionization fraction and the magnetic fields of the molecular clouds
(see, e.g., Ferri{\'e}re 2009; Croker et al. 2010), which is outside 
the scope of this paper. In a highly turbulent environment, the effect of a 
magnetic field on the cloud is to decrease the dissipation of kinetic 
energy, i.e., turbulent heating, and to lead to a lower gas kinetic temperature. 
Having already mentioned PDRs, we discuss in the following XDRs, CRDRs,
and the dissipation of turbulence.

\subsubsection{X-ray heating}\label{dis_xray}

In X-ray dominated regions (XDRs), X-rays photoionize atoms and molecules,
depositing a significant fraction of the primary and secondary electron energy
in heat (Maloney \etal\ 1996; Hollenbach \& Tielens 1999). Unlike UV photons,
hard X-ray photons are capable of deeply penetrating dense molecular
clouds and heating large amounts of gas. The X-ray heating rate is given by
\begin{eqnarray}
\Gamma_{\rm X}=1.2\times10^{-19}\,(\frac{n}{10^5\,{\rm cm^{-3}}})\,
(\frac{F_{\rm X}}{\rm erg\,cm^{-2}\,s^{-1}}) \nonumber\\ 
(\frac{N}{10^{22}\,{\rm cm^{-2}}})^{-0.9}\,
{\rm erg\,cm^{-3}\,s^{-1}}
\end{eqnarray}
(Maloney et al. 1996), where $n$ is gas density in units of cm$^{-3}$, $F_{\rm
X}$ is X-ray flux density in units of ${\rm erg\,cm^{-2}\,s^{-1}}$, and $N$
the column density of hydrogen attenuating the X-ray flux. For the dense cores
studied here, $N$ is of the order of $10^{23}\,{\rm cm^{-2}}$. In the region
associated with the Galactic center and encompassing a diameter of about 20
arcminutes, the total X-ray luminosity in the range 2 to 40 KeV (Muno
\etal\,2004; Koyama \etal\,2007; Yuasa \etal\,2008; Dogiel \etal\,2010) is
6$\times$10$^{36}$ ${\rm erg\,s^{-1}}$, yielding an X-ray flux density of
2.0$\times$10$^{-3}$ ${\rm erg\,cm^{-2}\,s^{-1}}$.

The gas cooling via gas-dust interaction can be expressed as 
\begin{equation} 
\Lambda_{\rm g-d}\,\sim\,4\times10^{-33}\,n^2\,T_{\rm
kin}^{1/2}(T_{\rm kin}-T_{\rm d})\,{\rm erg~cm^{-3}~s^{-1}}
\end{equation}
(Tielens 2005), where $T_{\rm kin}$ is the gas kinetic temperature and $T_{\rm
d}$ dust temperature. For a gas density range of 10$^4$ to
10$^6$\,cm$^{-3}$, accounting for the velocity gradient,
${\rm d}v/{\rm d}r$, the line cooling is approximated well by the
expression
\begin{equation} 
\Lambda_{\rm gas}\,\sim\,6\times10^{-29}\,n^{1/2}\,T_{\rm kin}^3\,{\rm d}v/{\rm
d}r\,{\rm erg~cm^{-3}~s^{-1}} 
\end{equation}
(Goldsmith \& Langer 1978; Goldsmith 2001; Papadopoulos 2010), where ${\rm
d}v/{\rm d}r$ is the velocity gradient adopted to
calculate the line cooling rate in the LVG radiative transfer models. 
The line cooling is through rotational lines of CO, its isotopologues $^{13}$CO
and C$^{18}$O, and other species (Goldsmith 2001). If the water is highly
abundant in the warm clouds of the CMZ, the line cooling from the water
emission will be important, and this will lead to lower gas kinetic temperatures
in this section. However, this will not drastically change our results,
because water is unlikely to be that dominant. Furthermore, water vapor only
affects the cooling rate and not the heating process, so that our evaluation of
relative efficiencies of different heating processes is not seriously
affected.

If X-ray heating dominates the heating process, the gas kinetic temperature 
can be estimated from the thermal equilibrium
\begin{equation}
{\bar\Gamma_{\rm X}}=\Lambda_{\rm g-d}+\Lambda_{\rm gas}\,.
\end{equation}
To simply solve the equation above, we set the dust temperature to $T_{\rm d}$=0\,K,
yielding a minimum $T_{\rm kin}$ value and its simple analytic solution as below
\begin{eqnarray} 
T_{\rm kin}=(\frac{(16\times10^{-8}\,n^3+720\,n^{0.5}\,{\rm d}v/{\rm d}r\,
F_{\rm X}\,(N/10^{23})^{-0.9})^{0.5}}{12\,{\rm d}v/{\rm d}r}\nonumber\\
-\frac{4\times10^{-4}\,n^{1.5}}{12\,{\rm d}v/{\rm d}r})^{2/3}\,.
\end{eqnarray}
Choosing a velocity gradient of 5\,\kms\,pc$^{-1}$, and adopting the observed
X-ray flux density of 2.0$\times$10$^{-3}$ ${\rm erg\,cm^{-2}\,s^{-1}}$, the
derived gas kinetic temperature is only 1\,K. It indicates that the observed X-ray
flux density cannot explain the high gas temperatures observed by formaldehyde.
Peculiar conditions in the CMZ, like a potentially high water vapor
abundance, enhancing cooling, or a high magnetic field inhibiting the dissipation
of turbulent motion, would not help to diminish the resulting discrepancy.

To simplify this: If the X-ray flux density were about 500 times higher than 
the observed value and line cooling dominates the cooling process, we could
derive an approximate gas temperature with
\begin{eqnarray}
T_{\rm kin}=45\,(\frac{n}{10^{4.5}\,{\rm cm}^{-3}})^{1/6}\,(\frac{{\rm d}v/{\rm
d}r}{5\,\kms\,{\rm pc}^{-1}})^{-1/3} \nonumber\\
(\frac{F_{\rm X}}{500\times2.0\times10^{-3}\,
{\rm erg\,cm^{-2}\,s^{-1}}})^{1/3}\,(\frac{N}{10^{23}\,{\rm cm^{-2}}})^{-0.3}\,{\rm K}\,.
\end{eqnarray}
This equation shows that $T_{\rm kin}$ depends weakly on the gas density, velocity
gradient, column density of hydrogen attenuating the X-ray emission, and X-ray
flux density. The first three parameters cannot change significantly. If
X-rays really play an important role in heating the gas, an X-ray flux density,
about three orders of magnitude higher than observed, is required. Such an
intense X-ray radiation field may exist in some spatially confined regions.
However, it cannot explain the high gas temperatures on the large spatial
scales of the GC. Recent observations (e.g., Eckart et al. 2012; Nowak et
al. 2012) confirm that the X-ray emission from Sgr~A$^*$ shows flares, almost
daily, by factors of a few to ten times over the quiescent emission level, and
rarely even up to more than 100 times that level on time scales from a few
minutes to a few hours. However, this is still much less than what is required
to explain our observed kinetic gas temperatures.

\subsubsection{Cosmic-ray heating}\label{dis_cr}

For the UV-shielded and mostly subsonic dense gas cores (e.g., dark clouds),
cosmic-ray heating is the major heating process, and it may also play an
important role in heating the gas in the GC because the cosmic-ray flux density
is enhanced by the supernovae in this region. The cosmic-ray heating rate is
given by 
\begin{equation} 
\Gamma_{\rm CR}\,\sim\,3.2\times10^{-28}\,n\,(\frac{\zeta_{\rm
CR}}{10^{-17}\,s^{-1}})\,{\rm erg~cm^{-3}~s^{-1}} 
\end{equation} 
(Goldsmith \& Langer 1978), where $\zeta_{\rm CR}$ is the total
cosmic-ray ionization rate. If the gas heating in the GC is dominated by the
cosmic-ray heating, we can obtain the gas kinetic temperature from the energy
balance equation 
\begin{equation} 
\Gamma_{\rm CR}=\Lambda_{\rm g-d}+\Lambda_{\rm gas}\,.
\end{equation} 
As above in $\S$~\ref{dis_xray}, we set the dust temperature to $T_{\rm d}$=0\,K,
yielding a minimum $T_{\rm kin}$ value and the following analytic solution as below
\begin{eqnarray} 
T_{\rm kin}=(\frac{(16\times10^{-8}\,n^3+768\,n^{0.5}\,{\rm d}v/{\rm
d}r\,\frac{\zeta_{\rm CR}}{10^{-17}\,s^{-1}})^{0.5}}{12\,{\rm
d}v/{\rm d}r}\nonumber\\
-\frac{4\times10^{-4}\,n^{1.5}}{12\,{\rm
d}v/{\rm d}r})^{2/3}\,.
\end{eqnarray}

Similarly, we can obtain a simple solution as below if line cooling dominates
the cooling process
\begin{eqnarray} 
T_{\rm kin}=6\,(\frac{n}{10^{4.5}\,{\rm cm}^{-3}})^{1/6}\,(\frac{{\rm d}v/{\rm
d}r}{5\,\kms\,{\rm pc}^{-1}})^{-1/3} \nonumber\\(\frac{\zeta_{\rm CR}}{10^{-17}\,{\rm
s}^{-1}})^{1/3}\,{\rm K}\,.
\end{eqnarray}
If cosmic-rays are the only heating source, gas temperatures are constrained by
three parameters: gas density, velocity gradient, and the cosmic-ray ionization
rate. It is reasonable to adopt a gas density of 10$^{4.5}$ cm$^{-3}$ (still low
enough to yield (almost) optically thin H$_2$CO 3$_{03}\rightarrow2_{02}$
emission) and a velocity gradient of 5\,\kmspc\, for the clouds in the GC. The
poorly determined parameter is the comic-ray ionization rate. Using the assumed
parameters, the gas kinetic temperatures from Equation~(14) are 22, 54, 70, and
122\,K for cosmic-ray ionization rates of 10$^{-15}$, 10$^{-14}$,
2$\times$10$^{-14}$, and 10$^{-13}$\,s$^{-1}$. Thus, if cosmic rays play an
important role in heating the gas, a cosmic-ray ionization rate of at least
1$-$2$\times$10$^{-14}$\,s$^{-1}$ is required to explain the observed
temperatures in the GC, which is about three orders of magnitude higher than in
the solar neighborhood (e.g., Farquhar et al. 1994). Such an enhanced flux of
cosmic-ray electrons is inferred in Sgr\,B2 by Yusef-Zadeh \etal\,(2007), and
is interpreted as the main molecular-gas heating source in this region.
The required high cosmic ray flux of 1$-$2$\times$10$^{-14}$\,s$^{-1}$
would lead to an H{\sc i} density of about 500~cm$^{-3}$ (G{\"u}sten et al. 1981).
Such a large H{\sc i} abundance will cause noticeable 21 cm signals, which have
indeed been seen in H{\sc i} absorption surveys toward some GC clouds (e.g.,
Schwarz et al.  1977; Lang et al. 2010).  For comparison, Bradford \etal\,
(2003) estimate that a high supernova rate in the nucleus of NGC\,253 results
in a cosmic-ray ionization rate of 1.5$-$5.3$\times$10$^{-14}$\,s$^{-1}$.  This
mechanism may also play an important role in regulating the gas in
ultraluminous infrared galaxies (ULIRGs) where the cosmic ray energy density
may be as high as 1000 times that of the local Galactic value or even
higher (Papadopoulos 2010; Papadopoulos \etal\,2011).

\subsubsection{Turbulent heating}\label{dis_turb}
The dissipation of turbulent kinetic energy provides a potentially important
heating source in Galactic astrophysical environments, such as interstellar
clouds (e.g., Falgarone \& Puget 1995) and the warm ionized medium (e.g.,
Minter \& Balser 1997), and in extragalactic environments, such as
intracluster cooling flows (Dennis \& Chandran 2005).
The observed large velocity dispersion in the GC requires energy input to
support the turbulence because the dynamic timescale is rather short, around
10$^6$ years. This implies a high turbulent heating rate.  Following Pan \&
Padoan (2009), the turbulent heating rate is given by
\begin{equation}
\Gamma_{\rm turb}=n\,\mu\,m_{\rm H}\bar\epsilon\,,
\end{equation}
where $n$ and $m_{\rm H}$ are the number density and the mass of the hydrogen
atom, $\mu$ is the mean molecular weight, $\mu$=2.35 for molecular clouds, and
$\bar\epsilon$ is the average dissipation rate per unit mass.  The average
dissipation rate per unit mass, is given by
\begin{equation}
\bar\epsilon=0.5\,\sigma_{\rm v}^3/L\,,
\end{equation}
where $\sigma_{\rm v}$ is the one-dimensional velocity dispersion and $L$
the size of the cloud. Replacing the expression for $\bar\epsilon$ in
equation~(16) and substituting units, the average turbulent heating rate is
\begin{equation} 
{\bar\Gamma_{\rm turb}}=3.3\times10^{-27}\,n\,\sigma_{\rm v}^3\,L^{-1}\,{\rm
erg~cm^{-3}~s^{-1}}\,,
\end{equation}
where the gas density $n$ is in units of cm$^{-3}$, the one-dimensional
velocity dispersion $\sigma_{\rm v}$ is in units of \kms, and the cloud size
$L$ is in units of pc. We can relate the one-dimensional velocity dispersion
and the observed FWHM line widths by the conversion $\sigma_{\rm v}$=$V_{\rm
FWHM}$/2.355 (Pan \& Padoan 2009). 

If turbulent heating dominates the heating process, the gas kinetic temperature
can be estimated from thermal equilibrium
\begin{equation}
{\bar\Gamma_{\rm turb}}=\Lambda_{\rm g-d}+\Lambda_{\rm gas}\,.
\end{equation}
To solve the equation above simply, we set the dust temperature to $T_{\rm d}$=0\,K,
yielding a minimum $T_{\rm kin}$ value and its simple analytic solution as
\begin{eqnarray} 
T_{\rm kin}=(\frac{(16\times10^{-8}\,n^3+7920\,n^{0.5}\,{\rm d}v/{\rm d}r\,\sigma_{\rm
v}^3\,L^{-1})^{0.5}}{12\,{\rm d}v/{\rm d}r}\nonumber\\
-\frac{4\times10^{-4}\,n^{1.5}}{12\,{\rm d}v/{\rm d}r})^{2/3}\,.
\end{eqnarray}\label{tkin_cal}
Choosing a velocity gradient of 5\,\kms\,pc$^{-1}$, an observed FWHM line
width of 20\kms, and a cloud size of 5\,pc, the derived gas kinetic temperature
ranges between 51 and 62\,K for a range in gas density of 10$^{4}$ to
10$^{5}$\,cm$^{-3}$.

\begin{figure}[t]
\vspace{-0.0cm}
\centering
\includegraphics[angle=0,width=0.5\textwidth]{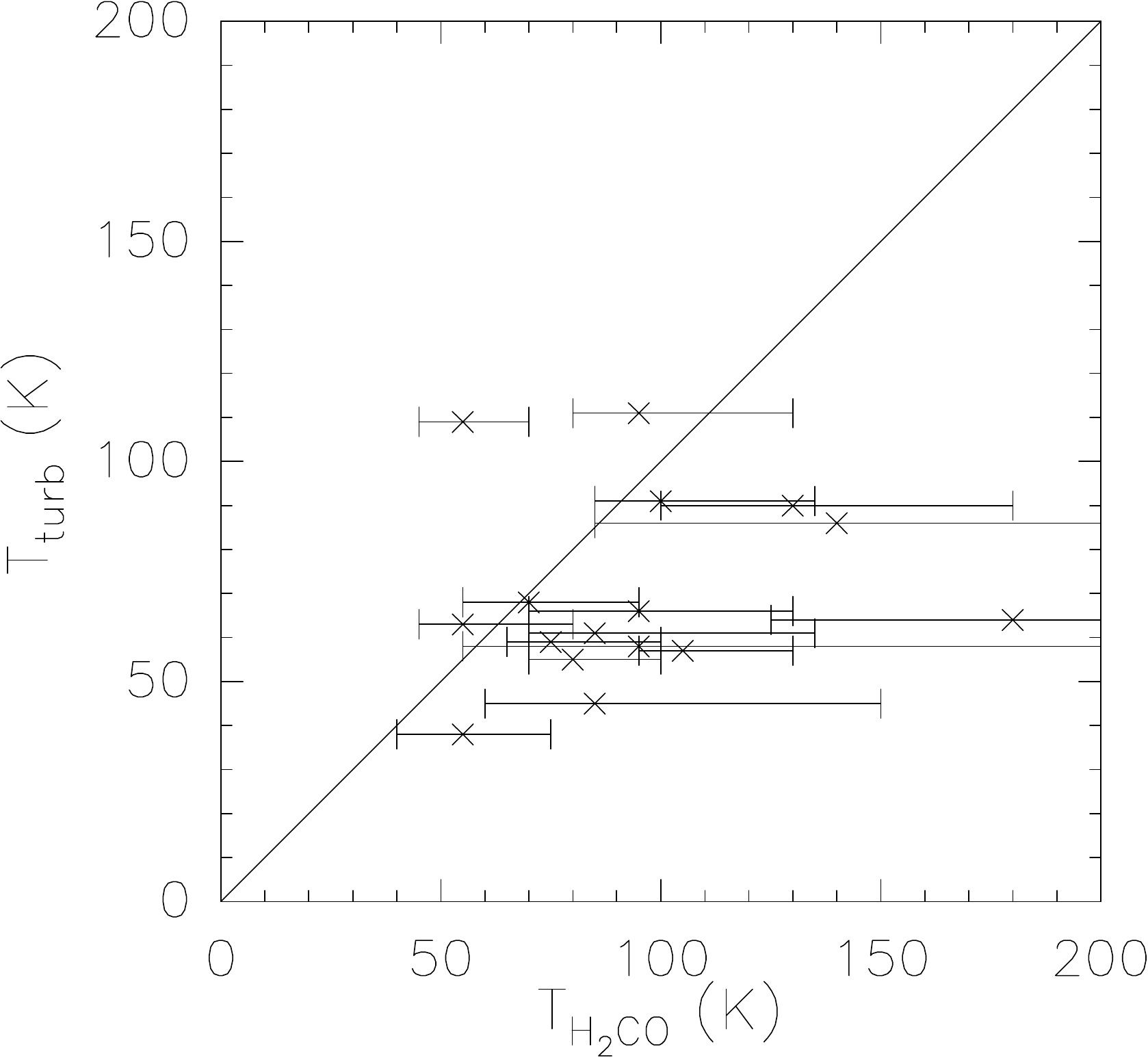}
\vspace{-0.0cm}
\caption{Gas temperatures estimated by turbulent heating versus those derived
from the H$_2$CO LVG models with [para-H$_2$CO] = 
10$^{-10}$. The solid line shows the relationship for the cases where both temperatures 
are the same.}
\label{tkin} 
\end{figure}

\begin{figure}[b]
\vspace{-0.0cm}
\centering
\includegraphics[angle=90,width=0.6\textwidth]{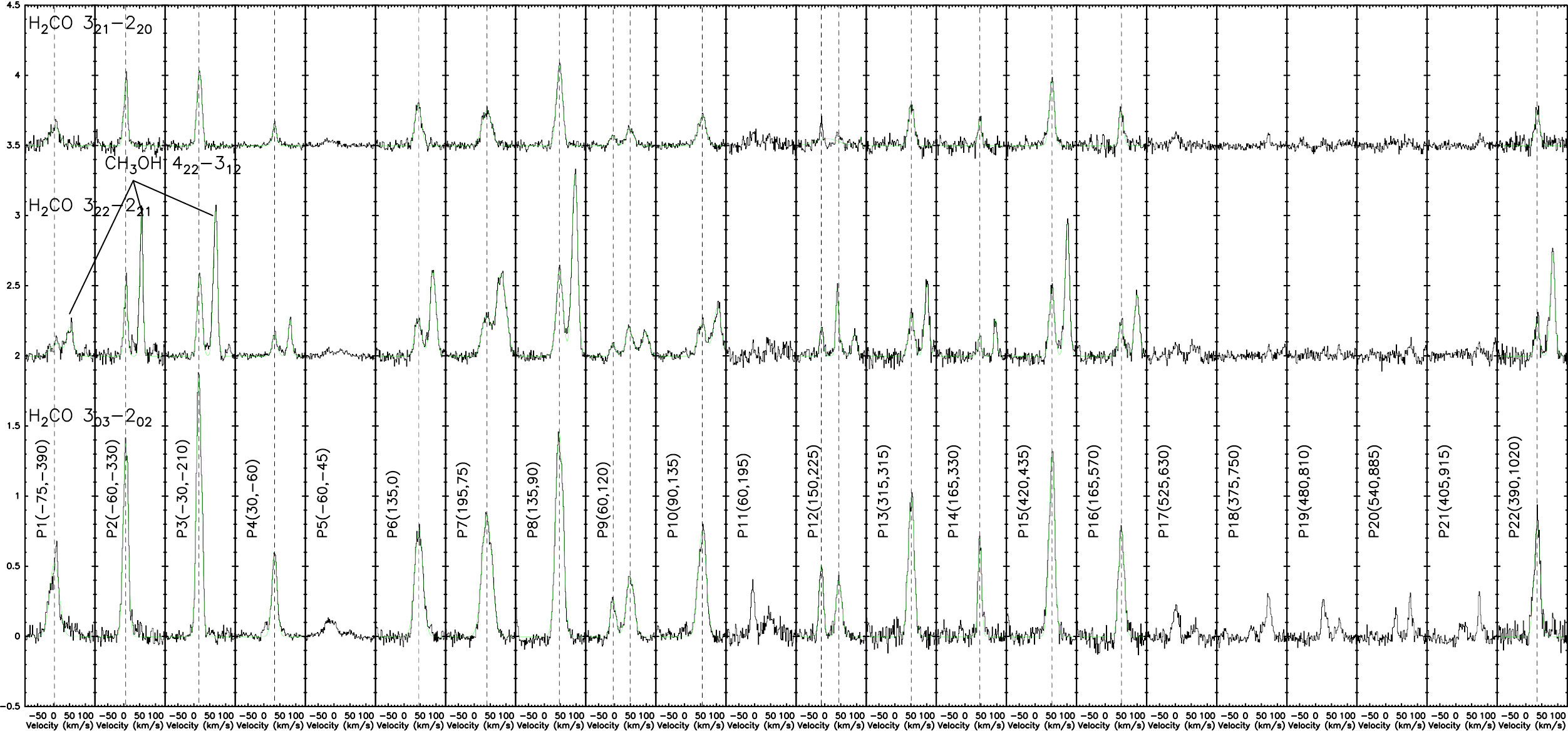}
\vspace{-0.0cm}
\caption{Spectra of different positions marked by the numbers in
Fig.~\ref{h2co303}. All transitions from a given position are presented in the
same panel but with different offsets along the y-axis on a $T_{\rm A}^*$ scale.
The Gaussian fits are indicated with green lines, and the central velocities
of the components are shown as dashed vertical lines.}
\label{spectra} 
\end{figure}

For gas densities $n$\,$\leq$\,10$^{5}$\,cm$^{-3}$ and the assumed parameters as
above, line cooling dominates the cooling process and the solution can be
simplified to
\begin{eqnarray} 
T_{\rm kin}=62\,(\frac{n}{10^{4.5}\,{\rm cm}^{-3}})^{1/6}\,(\frac{L}{5\,{\rm
pc}}\,\frac{{\rm d}v/{\rm d}r}{5\,\kms\,{\rm pc}^{-1}})^{-1/3} \nonumber\\(\frac{V_{\rm
FWHM}}{20\,\kms})\,{\rm K}\,.
\end{eqnarray} 
This equation underestimates the gas temperature by less than 10\% in
comparison with Equation~(20), and shows that the temperature depends only weakly
on the gas density, cloud size, and velocity gradient, but strongly
depends on line width. Considering the line widths in different clouds, we
can use Equation~(20) to calculate the temperatures and present the results in
the last column of Table~\ref{table2}. To compare the calculated temperatures
by turbulent heating with the ones derived from the H$_2$CO measurements in
$\S$~\ref{dis_lvg}, we plot both temperatures in Fig.~\ref{tkin}. In general,
the temperatures agree within the uncertainties, supporting turbulent heating
as a good candidate to heat the gas to the high temperatures observed in the
GC. High gas kinetic temperatures were also deduced from NH$_3$ absorption
lines toward Sgr B2 by Wilson et al. (1982). They suggested that the high gas
temperatures were caused by turbulence maintained by shearing forces as a
consequence of galactic rotation.
 
For turbulent and cosmic-ray heating, we cannot distinguish which
mechanism dominates the heating of molecular clouds in the GC. 
With a large interferometer such as ALMA, one can try to search for molecular
clumps with thermal line widths, i.e. with line widths that are dominated by
thermal motion. If such objects can be found, turbulent heating can be excluded
because the narrow line widths cannot be explained by turbulent heating, and the
cosmic-ray heating will then be the dominant process to heat the gas to high
temperatures. Future observations of $x(e)\,=\,\frac{n_e}{2n_{H_2}}$ (the
average ionization fraction) can also help distinguish between these two
heating mechanisms, because high cosmic-ray energy densities will boost this fraction,
unlike turbulence (Papadopoulos 2010 and references therein). 

The special thermal state of the GC clouds may be the average state of the
molecular ISM in ULIRGs, with a direct impact on their stellar IMF
(Papadopoulos et al. 2011). The high temperatures of molecular clouds on
large scales in the GC region may be driven by turbulent energy dissipation and/or
cosmic-rays instead of photons. Such a non-photon-driven thermal state of the
molecular gas provides an excellent template for studying the intial conditions
and star formation for the galaxy-sized gas in ULIRGs.

\section{Conclusions}
The $J_{\rm K_aK_c}$ = 3$_{03}\rightarrow2_{02}$, 3$_{22}\rightarrow2_{21}$,
and 3$_{21}\rightarrow2_{20}$ transitions of para-H$_2$CO were observed
simultaneously with the APEX telescope, covering an area of roughly
40$'$$\times$8$'$ along the Galactic plane with a linear resolution of 1.2\,pc,
including the Galactic center. The main results from these measurements follow. 

(1) The strongest line of the 218\,GHz H$_2$CO triplet, H$_2$CO
(3$_{03}\rightarrow2_{02}$), is widespread in the mapped region, and its
emission shows a morphology similar to ammonia (G{\"u}sten et al. 1981).

(2) The para-H$_2$CO abundance is found to be 0.5$-$1.2$\times$10$^{-9}$,
which is consistent with previous studies of formaldehyde absorption lines
at cm-wavelengths in the Galactic center region.

(3) Using LVG models, we can constrain gas kinetic temperatures to be of about
85~K for the Galactic center clouds, ranging from 50 to values above 100~K. 
While a systematic trend of (decreasing) kinetic temperature versus
(increasing) angular distance from the nucleus is not found, the clouds with
highest temperature ($T_{\rm kin}$ $>$ 100\,K) are all located near the center.
Molecular gas outside of the dense cores in the Galactic center region
is characterized by a gas kinetic temperature of 65\ppm10~K.

(4) The high temperatures found in the Galactic center region may be caused by
turbulent heating and/or cosmic-ray heating. Turbulent heating can readily heat
the gas to the values deduced from H$_2$CO. If cosmic-ray heating dominates the
heating process, a cosmic-ray ionization rate of at least
1$-$2$\times$10$^{-14}$ is required to explain the observed temperatures.
The high temperatures of molecular clouds on large scales in
the Galactic center region may be driven by turbulent energy dissipation and/or
cosmic-rays instead of photons. Such a non-photon-driven thermal state of the
molecular gas make such clouds excellent templates for the
starbursts found in ultraluminous infrared galaxies.


\begin{acknowledgements}
We thank the anonymous referee and the Editor Malcolm Walmsley for valuable
comments that improved this manuscript. We wish to thank Padelis Papadopoulos
for useful discussions. Y.A. acknowledges the supports by the grant 11003044
from the National Natural Science Foundation of China, and 2009's President
Excellent Thesis Award of the Chinese Academy of Sciences. This research has
made use of NASA’s Astrophysical Data System (ADS).

\end{acknowledgements}



\clearpage
\onecolumn

\begin{center}
\begin{table*}
\caption[]{H$_2$CO LVG results}\label{table2}
\begin{flushleft}
\begin{tabular}{lccc}
\hline
          & \multicolumn{1}{c}{[para-H$_2$CO]\,=\,10$^{-9}$} &  \multicolumn{1}{c}{[para-H$_2$CO]\,=\,10$^{-10}$}  & \\
          &  $T_{\rm kin}$ & $T_{\rm kin}$ & $T_{\rm turb}^a$ \\
Position  &  (K) &  (K) & (K) \\
\hline
P1  (--75'', --390'')            & 50$^{+10}_{-5}$ &  55$^{+15}_{-10}$          	& 107    \\
P2  (--69'', --330'')            & 75$^{+10}_{-10}$ &  80$^{+20}_{-10}$         	&  55   \\
P3  (--30'', --210'')            & 85$^{+10}_{-5}$ &  105$^{+25}_{-10}$ 		&  57  \\
P4  (+30'', --60'')              & 75$^{+20}_{-15}$ &   75$^{+25}_{-10}$ 		&  59  \\
P6  (+135'', 0'')                & 90$^{+25}_{-10}$ &   100$^{+35}_{-15}$ 		&  91   \\
P7  (+195'', 75'')               & 85$^{+15}_{-10}$ &  95$^{+35}_{-15}$ 		& 111   \\
P8  (+135'', +90'')              & 190$^{+60}_{-30}$ &  $>$250 			& 82   \\
P9 (+60''. +120''/ 49~\kms)      & 125$^{+95}_{-45}$ &  140$^{+120}_{-55}$ & 86  \\
P9 (+60'', +120''/ $-$4\kms)     & 90$^{+90}_{-25}$ &  95$^{+120}_{-40}$ 	&  58   \\
P10 (+90'', +135'')              & 110$^{+25}_{-20}$ & 130$^{+50}_{-30}$ 		& 90   \\
P12 (+150'', +225''/ $-$11\kms)  & 80$^{+50}_{-20}$ &   85$^{+65}_{-25}$ &  45   \\
P12 (+150'', +225''/ 44\kms)     & 55$^{+20}_{-10}$ &  55$^{+25}_{-10}$ 	&  63   \\
P13 (+315''. +315'')             & 85$^{+20}_{-15}$ &   95$^{+35}_{-25}$ 		&  66  \\
P14 (+165'', +330'')             & 55$^{+20}_{-10}$ &  55$^{+20}_{-15}$ 		&  38  \\
P15 (+420'', +435'')             & 125$^{+25}_{-25}$ &  180$^{+70}_{-55}$ 		&  64  \\
P16 (+165'', +570'')             & 80$^{+35}_{-15}$ & 85$^{+50}_{-15}$ 		&  61  \\
P22 (+390'', +1020'')            & 65$^{+20}_{-10}$ &  70$^{+25}_{-15}$ 		&  68  \\

\hline
\end{tabular}
\end{flushleft}
A velocity gradient of 5\kmspc, a filling factor of unity, and the H$_2$CO 
3$_{22}\rightarrow2_{21}$/3$_{03}\rightarrow2_{02}$ line ratios averaged
over 30$\arcsec$ boxes were adopted
in the LVG models to derive the solutions presented in Column~2$-$3. \\
\noindent $^a$ The temperatures in Column~4 are calculated with Equation~(20)
with a velocity gradient of 5\kmspc, a typical gas density of 10$^{4.5}$~cm$^{-3}$,
and a cloud size, i.e., turbulent scale, of 5~pc. 
\end{table*}
\end{center}

\clearpage

\appendix
\section{H$_2$CO line parameters}

\begin{center}
\begin{longtable}{llllllll}
\caption[H$_2$CO LVG results]{Line parameters}\label{table1}\\
\hline \hline
 & \multicolumn{2}{c}{Position} & & $T_{\rm mb}$  &  $V_{\rm LSR}$ & $\Delta V_{\rm 1/2}$   & $\int T_{\rm mb} dv$ \\
 & \multicolumn{2}{c}{(arcsec, arcsec)} & & (K) & (\kms) & (\kms) & (K\,\kms) \\
\hline
\endfirsthead
\multicolumn{8}{c}{\tablename\ \thetable{} -- continued from previous page} \\
\hline \hline
 & \multicolumn{2}{c}{Position} & & $T_{\rm mb}$  &  $V_{\rm LSR}$ & $\Delta V_{\rm 1/2}$   & $\int T_{\rm mb} dv$ \\
 & \multicolumn{2}{c}{(arcsec, arcsec)} & & (K) & (\kms) & (\kms) & (K\,\kms) \\
\hline
\endhead

\hline \multicolumn{8}{c}{{Continued on next page}} \\
\endfoot

\hline \hline
\endlastfoot
 & & & H$_2$CO 3$_{03}\rightarrow2_{02}$  &   0.75(0.04) &    2.9(0.4) &   35.4(1.1) &   26.4(0.7) \\
P 1 &  -75 & -390 & H$_2$CO 3$_{22}\rightarrow2_{21}$  &   0.13(0.03) &    8.0(2.6) &   39.8(6.4) &    5.1(0.7) \\
 & & & H$_2$CO 3$_{21}\rightarrow2_{20}$  &   0.19(0.03) &    2.5(1.2) &   38.3(3.2) &    7.3(0.5) \\
\hline
 & & & H$_2$CO 3$_{03}\rightarrow2_{02}$  &   1.95(0.05) &    6.5(0.2) &   18.8(0.4) &   36.7(0.7) \\
P 2 &  -60 & -330 & H$_2$CO 3$_{22}\rightarrow2_{21}$  &   0.72(0.04) &    7.1(0.3) &   13.1(0.8) &    9.5(0.5) \\
 & & & H$_2$CO 3$_{21}\rightarrow2_{20}$  &   0.65(0.03) &    6.7(0.2) &   14.3(0.7) &    9.5(0.3) \\
\hline
 & & & H$_2$CO 3$_{03}\rightarrow2_{02}$  &   2.64(0.04) &   16.4(0.1) &   19.4(0.2) &   51.2(0.5) \\
P 3 &  -30 & -210 & H$_2$CO 3$_{22}\rightarrow2_{21}$  &   0.83(0.03) &   18.4(0.2) &   18.0(0.4) &   15.0(0.3) \\
 & & & H$_2$CO 3$_{21}\rightarrow2_{20}$  &   0.76(0.02) &   17.7(0.2) &   17.7(0.4) &   13.5(0.2) \\
\hline
 & & & H$_2$CO 3$_{03}\rightarrow2_{02}$  &   0.83(0.02) &   33.2(0.2) &   20.1(0.5) &   16.7(0.3) \\
P 4 &   30 &  -60 & H$_2$CO 3$_{22}\rightarrow2_{21}$  &   0.21(0.01) &   34.0(0.4) &   19.5(1.4) &    4.2(0.2) \\
 & & & H$_2$CO 3$_{21}\rightarrow2_{20}$  &   0.21(0.01) &   33.5(0.3) &   16.1(1.0) &    3.4(0.1) \\
\hline
 & & & H$_2$CO 3$_{03}\rightarrow2_{02}$  &   1.08(0.04) &   46.2(0.3) &   30.2(0.7) &   32.7(0.7) \\
P 6 &  135 &    0 & H$_2$CO 3$_{22}\rightarrow2_{21}$  &   0.37(0.03) &   44.6(0.6) &   26.0(1.3) &    9.6(0.4) \\
 & & & H$_2$CO 3$_{21}\rightarrow2_{20}$  &   0.41(0.02) &   44.2(0.4) &   24.4(0.9) &   10.2(0.3) \\
\hline
 & & & H$_2$CO 3$_{03}\rightarrow2_{02}$  &   1.23(0.04) &   40.2(0.3) &   36.8(0.6) &   45.0(0.7) \\
P 7 &  195 &   75 & H$_2$CO 3$_{22}\rightarrow2_{21}$  &   0.40(0.02) &   37.3(0.6) &   31.5(1.4) &   12.7(0.5) \\
 & & & H$_2$CO 3$_{21}\rightarrow2_{20}$  &   0.36(0.02) &   39.8(0.4) &   34.9(1.0) &   12.4(0.3) \\
\hline
 & & & H$_2$CO 3$_{03}\rightarrow2_{02}$  &   2.07(0.07) &   47.0(0.3) &   27.4(0.6) &   56.5(1.1) \\
P 8 &  135 &   90 & H$_2$CO 3$_{22}\rightarrow2_{21}$  &   0.85(0.03) &   48.2(0.2) &   26.0(0.6) &   22.2(0.4) \\
 & & & H$_2$CO 3$_{21}\rightarrow2_{20}$  &   0.81(0.02) &   47.5(0.2) &   23.0(0.4) &   18.6(0.3) \\
\hline
 & & & H$_2$CO 3$_{03}\rightarrow2_{02}$  &   0.60(0.02) &   49.3(0.3) &   28.6(0.6) &   17.3(0.3) \\
P 9 &   60 &  120 & H$_2$CO 3$_{22}\rightarrow2_{21}$  &   0.16(0.02) &   49.3(...)$^a$ &   38.2(4.6) &    5.9(0.8) \\
 & & & H$_2$CO 3$_{21}\rightarrow2_{20}$  &   0.17(0.01) &   48.5(0.7) &   28.9(1.7) &    4.8(0.2) \\
\hline
 & & & H$_2$CO 3$_{03}\rightarrow2_{02}$  &   0.37(0.02) &   -4.2(0.4) &   19.8(0.9) &    7.3(0.3) \\
P 9 &   60 &  120 & H$_2$CO 3$_{22}\rightarrow2_{21}$  &   0.12(0.02) &   -4.2(...)$^a$ &   16.9(2.3) &    2.1(0.3) \\
 & & & H$_2$CO 3$_{21}\rightarrow2_{20}$  &   0.11(0.01) &   -4.7(0.9) &   20.2(2.0) &    2.1(0.2) \\
\hline
 & & & H$_2$CO 3$_{03}\rightarrow2_{02}$  &   1.07(0.04) &   55.5(0.2) &   30.0(0.6) &   31.9(0.5) \\
P10 &   90 &  135 & H$_2$CO 3$_{22}\rightarrow2_{21}$  &   0.32(0.02) &   54.8(0.6) &   31.4(1.6) &   10.2(0.4) \\
 & & & H$_2$CO 3$_{21}\rightarrow2_{20}$  &   0.29(0.02) &   55.0(0.4) &   30.1(1.1) &    8.8(0.3) \\
\hline
 & & & H$_2$CO 3$_{03}\rightarrow2_{02}$  &   0.56(0.03) &   44.4(0.3) &   21.5(0.9) &   12.2(0.4) \\
P12 &  150 &  225 & H$_2$CO 3$_{22}\rightarrow2_{21}$  &   0.20(0.03) &   44.4(...)$^a$ &   18.3(2.4) &    3.8(0.5) \\
 & & & H$_2$CO 3$_{21}\rightarrow2_{20}$  &   0.12(0.02) &   42.4(1.3) &   19.8(2.9) &    2.3(0.3) \\
\hline
 & & & H$_2$CO 3$_{03}\rightarrow2_{02}$  &   0.72(0.03) &  -10.9(0.2) &   15.7(0.5) &   11.4(0.3) \\
P12 &  150 &  225 & H$_2$CO 3$_{22}\rightarrow2_{21}$  &   0.29(0.03) &  -10.9(...)$^a$ &   13.5(1.2) &    3.9(0.3) \\
 & & & H$_2$CO 3$_{21}\rightarrow2_{20}$  &   0.25(0.02) &  -10.6(0.4) &   11.4(1.3) &    2.9(0.3) \\
\hline
 & & & H$_2$CO 3$_{03}\rightarrow2_{02}$  &   1.41(0.05) &   51.3(0.2) &   22.3(0.5) &   31.5(0.7) \\
P13 &  315 &  315 & H$_2$CO 3$_{22}\rightarrow2_{21}$  &   0.41(0.04) &   52.2(0.7) &   21.1(1.8) &    8.7(0.6) \\
 & & & H$_2$CO 3$_{21}\rightarrow2_{20}$  &   0.40(0.03) &   51.5(0.5) &   21.5(1.2) &    8.7(0.4) \\
\hline
 & & & H$_2$CO 3$_{03}\rightarrow2_{02}$  &   0.99(0.04) &   46.7(0.2) &   13.6(0.5) &   13.5(0.4) \\
P14 &  165 &  330 & H$_2$CO 3$_{22}\rightarrow2_{21}$  &   0.20(0.03) &   45.9(1.0) &   12.5(3.2) &    2.5(0.4) \\
 & & & H$_2$CO 3$_{21}\rightarrow2_{20}$  &   0.28(0.03) &   46.7(0.5) &   12.7(1.2) &    3.6(0.3) \\
\hline
 & & & H$_2$CO 3$_{03}\rightarrow2_{02}$  &   1.83(0.06) &   52.9(0.2) &   21.8(0.5) &   39.9(0.7) \\
P15 &  420 &  435 & H$_2$CO 3$_{22}\rightarrow2_{21}$  &   0.68(0.04) &   52.7(0.3) &   19.6(0.8) &   13.4(0.5) \\
 & & & H$_2$CO 3$_{21}\rightarrow2_{20}$  &   0.65(0.02) &   52.5(0.2) &   19.5(0.6) &   12.8(0.3) \\
\hline
 & & & H$_2$CO 3$_{03}\rightarrow2_{02}$  &   1.09(0.05) &   51.0(0.3) &   20.9(0.7) &   22.8(0.7) \\
P16 &  165 &  570 & H$_2$CO 3$_{22}\rightarrow2_{21}$  &   0.33(0.04) &   50.7(0.7) &   18.9(2.1) &    6.3(0.5) \\
 & & & H$_2$CO 3$_{21}\rightarrow2_{20}$  &   0.36(0.04) &   50.3(0.6) &   16.5(1.4) &    5.9(0.4) \\
\hline
 & & & H$_2$CO 3$_{03}\rightarrow2_{02}$  &   1.15(0.06) &   35.7(0.3) &   22.9(0.8) &   26.1(0.8) \\
P22 &  390 & 1020 & H$_2$CO 3$_{22}\rightarrow2_{21}$  &   0.37(0.04) &   35.4(0.6) &   16.2(1.7) &    6.2(0.5) \\
 & & & H$_2$CO 3$_{21}\rightarrow2_{20}$  &   0.36(0.04) &   35.8(0.6) &   18.2(1.5) &    6.7(0.5) \\
\hline

\end{longtable}
\begin{flushleft}
\noindent $^a$ To obtain Gaussian fits of the H$_2$CO 3$_{22}\rightarrow2_{21}$ transition
at locations with two velocity components, we adopt the same central velocities
as determined for the H$_2$CO 3$_{03}\rightarrow2_{02}$ transition. 
\end{flushleft}

\end{center}

\section{Online material}
\subsection{H$_2$CO spectral lines}

\clearpage
\subsection{Velocity channel maps}

\begin{figure}[b]
\vspace{-0.0cm}
\centering
\includegraphics[angle=0,width=1.0\textwidth]{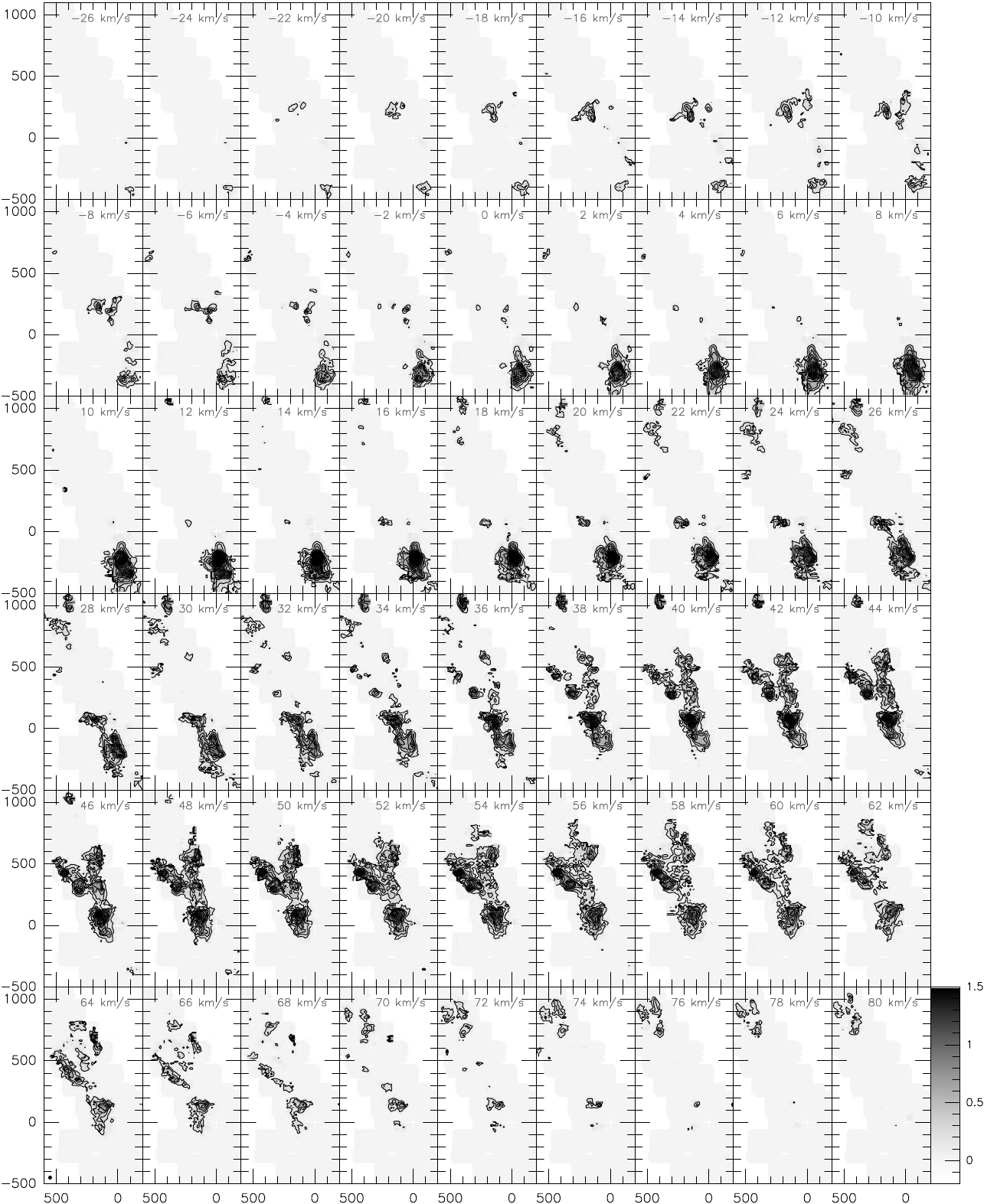}
\vspace{-0.0cm}
\caption{Channel maps of the H$_2$CO 3$_{03}-2_{02}$ emission. Black contour
levels for the molecular line emission (on a $T_{\rm A}^*$ scale) are from 0.16~K
(2$\sigma$) in s.pdf of 0.16~K. Velocity channels range from $-$27 to $+$81~\kms\, in
s.pdf of 2~\kms. The central velocities of the channel maps are shown at the
top of each panel. The wedge at the side shows the intensity
scale of the line emission. The beam size of 30$\arcsec$ is shown in the
bottom-left corner.}
\label{h2co303_channel} 
\end{figure}

\begin{figure}[t]
\vspace{-0.0cm}
\centering
\includegraphics[angle=0,width=1.0\textwidth]{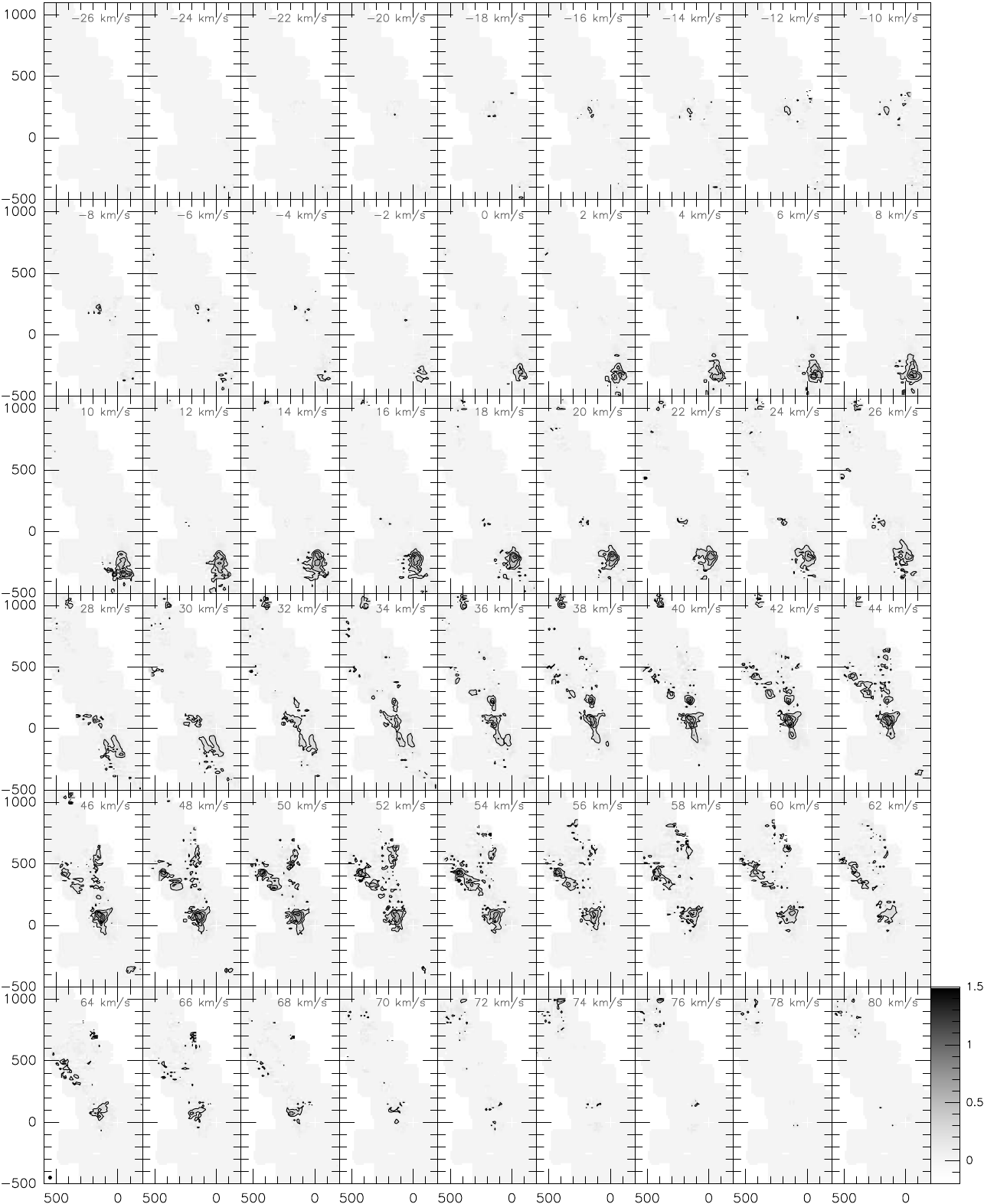}
\vspace{-0.0cm}
\caption{Same as Fig.~\ref{h2co303_channel} but for the H$_2$CO 3$_{22}-2_{21}$ emission.}
\label{h2co322_channel} 
\end{figure}

\begin{figure}[t]
\vspace{-0.0cm}
\centering
\includegraphics[angle=0,width=1.0\textwidth]{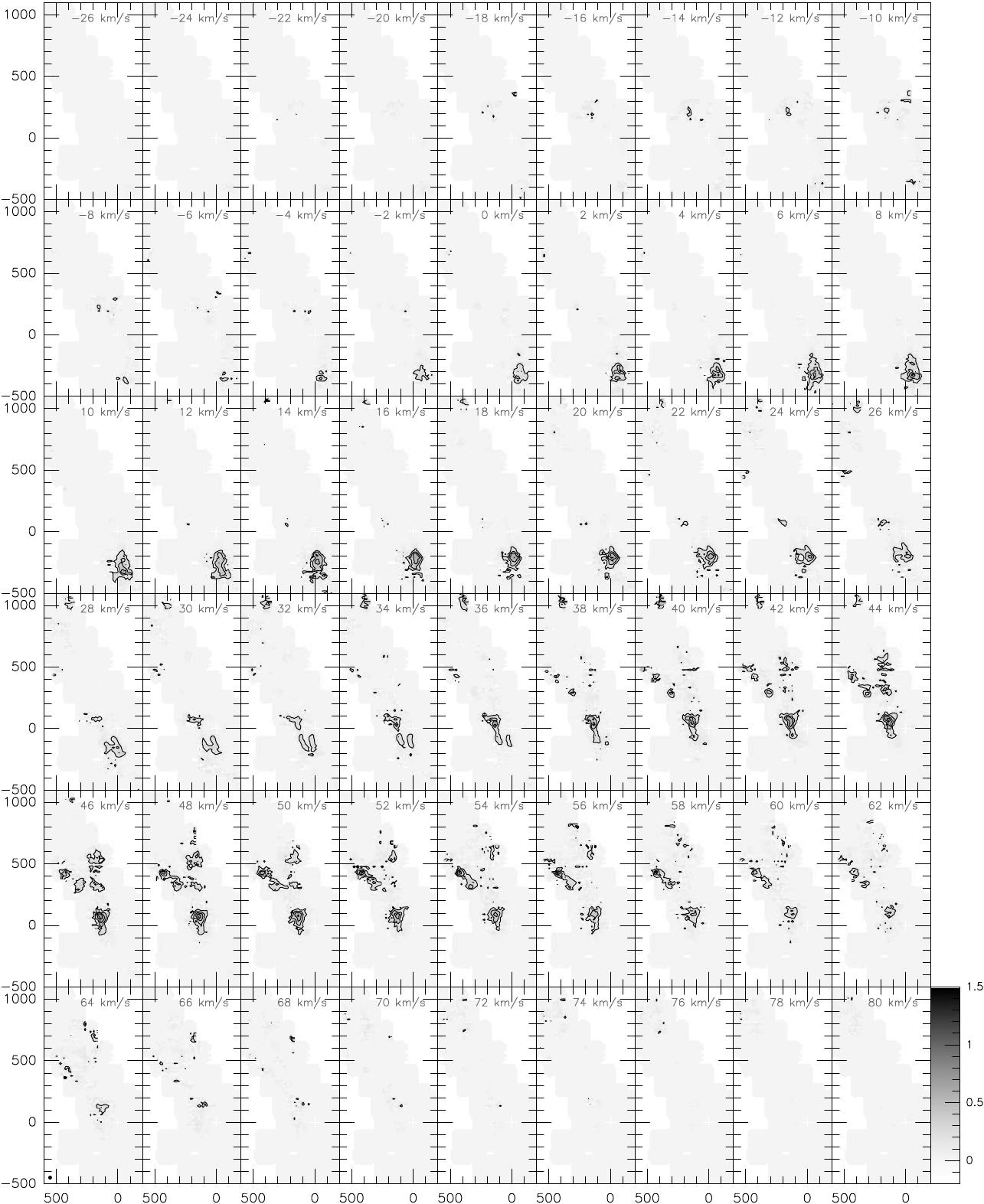}
\vspace{-0.0cm}
\caption{Same as Fig.~\ref{h2co303_channel} but for the H$_2$CO 3$_{21}-2_{20}$ emission.}
\label{h2co321_channel} 
\end{figure}

\end{document}